\documentclass[aps,prb,10pt,twocolumn,amsmath,amssymb,
    longbibliography,superscriptaddress,nofootinbib,
    showpacs,preprintnumbers]{revtex4-1}

\usepackage{epsfig}
\usepackage{array}
\usepackage{multirow}
\usepackage{amsmath}
\usepackage{graphicx}
\usepackage{bm,bbm}
\usepackage[colorlinks=true,citecolor=blue,linkcolor=blue]{hyperref}

\newcommand{\normalket}[1]{\vert #1 \rangle}

\newcommand{\ket}[1]{\normalket{#1}}

\newcommand{\normalbraket}[2]{\langle #1 \vert #2 \rangle}

\newcommand{\braket}[2]{\normalbraket{#1}{#2}}
\newcommand{\normalmatel}[3]{\langle #1 \vert #2 \vert #3 \rangle}
\newcommand{\bigmatel}[3]{\left\langle #1 \middle\vert #2 \middle\vert #3 \right\rangle}
\newcommand{\matel}[3]{\normalmatel{#1}{#2}{#3}}

\def\doteq{\,\overset{\boldsymbol{.}}{=}\,}
\def\dotapprox{\,\overset{\boldsymbol{.}}{\approx}\,}

\begin{document}

\title{Computation of intrinsic spin Hall conductivities from first principles using maximally-localized Wannier functions}

\author{Ji Hoon Ryoo}
\author{Cheol-Hwan Park}
\email{cheolhwan@snu.ac.kr}
\affiliation{Department of Physics, Seoul National University, Seoul 08826, Korea}
\author{Ivo Souza}
\affiliation{Centro de F\'isica de Materiales, Universidad del Pa\'is Vasco (UPV/EHU), 20018 San Sebasti\'an, Spain}
\affiliation{Ikerbasque Foundation, 48013 Bilbao, Spain}
\date{\today}

\begin{abstract}

We present a method to compute the intrinsic spin Hall conductivity
from first principles using an interpolation scheme based on
maximally-localized Wannier functions. After obtaining the relevant
matrix elements
among the {\it ab initio} Bloch states calculated on a coarse
$k$-point mesh, we Fourier transform them to find the corresponding
matrix elements between
Wannier states.
We then perform an inverse Fourier transform to interpolate
the velocity and spin-current matrix elements 
onto a dense 
$k$-point 
mesh,
and use them to evaluate
the spin Hall conductivity as a Brillouin-zone integral.
This strategy has a much lower computational cost than
a direct {\it ab initio} calculation, without sacrificing the accuracy.
We demonstrate that the spin Hall conductivities of platinum and doped
gallium arsenide, computed with our interpolation scheme as a function
of the Fermi energy, are in good agreement
with those obtained in previous first-principles studies. 
We also discuss certain approximations that can be made,
in the spirit of the tight-binding method, to simplify the
calculation of the velocity and spin-current matrix elements in the Wannier
representation.

\end{abstract}

\maketitle

\section{introduction}
\label{sec:intro}

The spin Hall effect is a phenomenon in which a transverse spin current
is generated in response to a bias voltage.~\cite{sinova2015RMP,sinova2012NatMater, jungwirth2012NatMater}
Together with its inverse effect,\cite{Hirsch1999PRL, Saitoh2006APL}
the conversion of charge current into spin current
by the spin Hall effect has been used
to electrically inject or detect spin,\cite{Kimura2007PRL}
to fabricate spin field-effect transistors,\cite{WunderlichScience2010}
to manipulate spin dynamics in magnetic microstructures,\cite{Ando2008PRL}
and to exert significant spin torques in an adjacent magnetic film.\cite{Liu2011PRL,Liu2012Science}
In order to utilize the spin Hall effect, it is desirable to find materials with large spin Hall conductivities (SHCs).

In the weak-disorder limit, the SHC
of conducting systems such as metals and doped semiconductors
is usually divided into three
contributions:\cite{nagaosa2010RMP,sinova2015RMP} (i) the
skew-scattering contribution, which is inversely proportional to the
concentration of impurities; (ii) the intrinsic contribution, which is
independent of the concentration of impurities and whose value is
fully determined by the electronic properties of the pristine
material; and (iii) the side-jump contribution, which is independent
of the concentration of impurities but whose value depends on the
details of the disorder potential.

Contrary to the intrinsic contribution, the other two
  contributions depend on disorder, which is why they are considered
  extrinsic.
  Calculations of the extrinsic
  contributions to the SHC have been mostly limited to simple
  toy-model systems such as two-dimensional
  electron~\cite{inoue2004PRB} or hole~\cite{bernevig2005PRL}
  gases. More recently, a first-principles methodology for calculating
  those contributions has been developed.\cite{chadova2015PRB} It has
  been reported that in many materials with strong spin-orbit coupling
  the intrinsic contribution accounts for a significant portion of the
  SHC (see, for example, Ref.~\onlinecite{Guo2008PRL}; for more
  details, see Ref.~\onlinecite{nagaosa2010RMP,sinova2015RMP} and
  references therein).
For this reason, most first-principles studies
of the spin Hall effect in conducting systems
have focused exclusively on the intrinsic contribution.

The intrinsic SHC can be calculated from the Kubo formula
  given in Eq.~({\ref{eq:sigma}}) below.
  The needed ingredients are the Bloch eigenstates and energy
  eigenvalues of the pristine crystal, which are typically obtained
  from a
  first-principles calculation based on density-functional theory.
Such calculations were initially performed for simple
semiconductors~\cite{Guo2005PRL} and metals,\cite{Guo2008PRL,
  Guo2009JAP} and recently they have been carried out for more complex
systems such as metallic alloys.\cite{Sui2017PRB}
Those studies revealed that a rather dense $k$-point mesh is often
needed to converge the calculation. This happens when, in some regions
of the Brillouin zone (BZ), occupied and empty states are both present
close to the Fermi level, leading to a resonant enhancement of the
energy denominator in Eq.~({\ref{eq:sigma}}).
Calculating the needed Bloch states directly from first principles over a 
large number of $k$ points is very demanding, and it would be
desirable to develop more efficient algorithms.

In this work, we present a method that circumvents the need to
calculate from first principles the Bloch states on the dense
$k$-point mesh where Eq.~({\ref{eq:sigma}}) is to be evaluated.
This is achieved by constructing maximally-localized Wannier
functions~\cite{Marzari1997PRB,Souza2001PRB,mostofi2008,Marzari2012RMP}
(MLWFs)
from the output of a conventional first-principles calculation carried
out on a relatively coarse $k$-point mesh. The locality of the MLWFs
is then exploited to interpolate the needed matrix elements
across the BZ.
This ``Wannier interpolation'' strategy has been used previously to
evaluate related quantities, such as the
anomalous Hall conductivity and the optical
conductivity.\cite{Wang2006PRB,Yates2007PRB} It was found to
reduce very significantly the computational cost, while retaining the
full accuracy of a direct {\it ab initio} calculation.

Wannier interpolation has also been used in a few previous works to
evaluate the intrinsic SHC.~\cite{Feng2012PRB,Sui2017PRB,
Sun2016PRL,Zhang2017PRB,Zelezny2017PRL,derunova2019sciadv,Zhang2018NJP}
Although few details were provided, it appears that some simplifying
assumptions were made whose impact on the calculated SHC remains to be
assessed. In particular, it was implicitly assumed in those works that
the Bloch subspace spanned by the MLWFs is left invariant under the
action of the Pauli spin operator ({\it i.e.}, that the spin operator
has vanishing matrix elements between a pair of Bloch states one of
which lies inside that subspace, and the other lies outside),
or of the velocity operator.
Our
method does not rely on this assumption, allowing us to check its
validity. We also investigate the impact on the SHC of another
approximation that is often made, in the spirit of the tight-binding
method: to assume,
when calculating velocity and spin-current matrix elements,
that the Wannier functions are perfectly localized at discrete points in space.

The paper is organized as follows. In Sec.~\ref{sec:formalism}, we
describe the proposed methodology as follows. We begin in
Sec.~\ref{sec:kubo} by reviewing the Kubo formula for the intrinsic
SHC.  In Sec.~\ref{sec:interpol}, we describe
the Wannier-interpolation scheme for computing
the $k$-space matrix elements and eigenvalues
appearing in the Kubo formula.  In Sec.~\ref{sec:real-space}, we
explain how the corresponding real-space matrix elements between MLWFs
are evaluated. Finally, in Sec.~\ref{sec:approx} we discuss the two
approximations that have been made in previous works to simplify the
calculation of the SHC by Wannier interpolation. The computational
details of the illustrative calculations for platinum and doped
gallium arsenide are given in Sec.~\ref{sec:comp-details}, and
the results of those calculations are presented and discussed in
Sec.~\ref{sec:results}.
After discussing in Sec.~\ref{sec:comp_cost}
the computational advantages of making the two
approximations mentioned above,
we conclude in Sec.~\ref{sec:conclusion} with
a summary.

\section{Formalism}
\label{sec:formalism}

\subsection{Kubo formula for the intrinsic spin Hall conductivity}
\label{sec:kubo}

We work in the independent-particle approximation, and consider a
crystal described by a lattice-periodic Hamiltonian $H$. The Bloch
states $\ket{\psi_{n\mathbf{k}}}$ with band index $n$ and
wavevector $\mathbf{k}$ satisfy
$H\ket{\psi_{n\mathbf{k}}}=\varepsilon_{n{\bf k}}\ket{\psi_{n\mathbf{k}}}$.
We adopt a convention where, for any
lattice-periodic operator $O$, $O_\mathbf{k}$ denotes
$e^{-i\mathbf{k}\cdot \mathbf{r}}O e^{i\mathbf{k}\cdot \mathbf{r}}$. Introducing
the cell-periodic part
$\ket{u_{n\mathbf{k}}}
=e^{-i\mathbf{k}\cdot\mathbf{r}}\ket{\psi_{n\mathbf{k}}}$
of the Bloch states, we find that they satisfy
$H_{\bf k}\ket{u_{n\mathbf{k}}}=\varepsilon_{n{\bf k}}\ket{u_{n\mathbf{k}}}$.

The component $\sigma^z _{xy}$ of the SHC tensor describes a
spin current flowing along $x$ and polarized along $z$, induced at
linear order by an electric field pointing along~$y$.
The intrinsic contribution at frequency $\omega$ is given
by the Kubo formula~\cite{Guo2005PRL,Guo2008PRL}
\begin{widetext}
\begin{equation}
\sigma^z _{xy}(\omega)=
\frac{e}{\hbar}\sum_{\bf k} \sum_n \sum_{n'\neq n}
\left(f_{n{\bf k}}-f_{n'{\bf k}} \right)\,\cdot\,
\frac{{\rm{Im}}
\left[
\matel{u_{n{\bf k}}}{j^z_{x,\mathbf{k}}}{u_{n'{\bf k}}}
\matel{u_{n'{\bf k}}}{v_{y,\mathbf{k}}}{u_{n{\bf k}}}
\right]
}
{\left(\varepsilon_{n{\bf k}}-\varepsilon_{n'{\bf k}} \right)^2
-\left(\hbar\omega+i\delta\right)^2}\,,
\label{eq:sigma}
\end{equation}
\end{widetext}
where $f_{n{\bf k}}$ is the Fermi-Dirac occupation factor for energy
$\varepsilon_{n{\bf k}}$ at a given temperature $T$
and Fermi energy
$E_{\rm F}$, and $\delta$ is a positive
infinitesimal.
(Throughout this paper, we assume $T=0~\textrm{K}$.)
The velocity operator
is defined as
\begin{equation}
v_{y,{\bf k}} = \frac{1}{i\hbar} \left[ y,\,H_{\bf k} \right]=
\frac{1}{\hbar} \partial_y H_{\bf k}
    \label{eq:vy}
\end{equation}
where $\partial_y=\partial/\partial k_y$, and the spin-current operator as
\begin{equation}
j^z _{x,{\bf k}} =
\frac{1}{2} \{s^z,\,v_{x,{\bf k}} \}\,,
\label{eq:jzx}
\end{equation}
where $s^z=(\hbar/2)\sigma^z$
with $\sigma^z$ the Pauli matrix.

\subsection{Wannier-interpolation scheme}
\label{sec:interpol}

We now describe the interpolation scheme for evaluating, at an
arbitrary point ${\bf k}$, the energy eigenvalues
and matrix elements
appearing in Eq.~\eqref{eq:sigma}.
We assume that a set of $N_\textrm{W}$ MLWFs per unit cell has been
obtained in a post-processing step following a standard
first-principles
calculation.\cite{Marzari1997PRB,Souza2001PRB,mostofi2008,Marzari2012RMP}
By construction, those MLWFs correctly describe the {\it ab initio}
electronic states over some energy range spanning the valence and a
number of low-lying conduction bands. We denote by
$\left|{\bf R}n\right>$
the Wannier function centered at ${\bf R}+{\boldsymbol \tau}_n$, where
${\bf R}$ is a lattice vector and ${\boldsymbol \tau}_n$ is a Wannier
center in the home cell ${\bf R}={\bf 0}$, with $n$ running from $1$
to $N_\textrm{W}$. For each $n$, we define the Bloch-sum state
\begin{equation}
\left| u^{({\rm W})}_{n{\bf k}}\right>
=\sum_{\bf R}\, e^{-i{\bf k}\cdot \left({\bf r}-{\bf R}\right)} \left|{\bf R}n\right>\,,
\label{eq:BlochSum}
\end{equation}
where the subscript (W) stands for ``Wannier gauge.''  Because the
MLWFs are well localized in real space, these cell-periodic
states are smooth functions of ${\bf k}$.

\subsubsection{Energy eigenvalues}
\label{sec:eigvals}

We begin by reviewing the procedure for interpolating the band
energies.\cite{Souza2001PRB} Once the Hamiltonian matrix elements
$\left<{\bf 0}m\right| H \left|{\bf R}n\right>$ between the MLWFs have
been tabulated (see Sec.~\ref{sec:real-space}), the corresponding
matrix elements between the
Bloch-sum states can be evaluated at any given ${\bf k}$ as a Fourier
sum,
\begin{eqnarray}
H_{mn{\bf k}} ^{({\rm W})}&=&
\left< u^{({\rm W})}_{m{\bf k}}\right| H_{\bf k} \left| u^{({\rm W})}_{n{\bf k}}\right>
\nonumber\\
&=&\sum_{\bf R}\, e^{i{\bf k}\cdot {\bf R}}
\left<{\bf 0}m\right| H \left|{\bf R}n\right>\,.
\label{eq:Hmn}
\end{eqnarray}
The interpolated eigenvalues are then obtained by diagonalizing this
$N_{\rm W}\times N_{\rm W}$ matrix,
\begin{equation}
  [U_{\bf k}^\dagger H^{({\rm W})}_{\bf k}U_{\bf k}]_{mn} =
  H^{({\rm H})} _{mn{\bf k}} = \varepsilon^{({\rm H})}_{m{\bf k}}\, \delta_{mn}.
\label{eq:epsilon}
\end{equation}
Here $U_{\mathbf{k}}$ is a
unitary matrix of rank $N_{\rm W}$, and (H) stands for ``Hamiltonian
gauge.''  If the {\it ab initio} bands that were wannierized form an
isolated group, the eigenvalues $\varepsilon^{({\rm H})} _{m{\bf k}}$
interpolate between the first-principles eigenvalues
$\varepsilon_{m{\bf q}}$ on the Monkhorst-Pack grid $\{{\bf q}\}$ that
was used for constructing the MLWFs.  For a non-isolated
(``entangled'') group of bands, the interpolation is accurate only
within the inner
energy window used in the disentanglement step.\cite{Souza2001PRB}

Defining the interpolated Bloch eigenstates as
\begin{equation}
\left| u^{({\rm H})}_{n{\bf k}}\right>=
\sum_m \left| u^{({\rm W})}_{m{\bf k}}\right>\,U_{mn{\bf k}}
\label{eq:U}
\end{equation}
where the summation goes from $1$ to $N_{\rm W}$,
Eq.~\eqref{eq:epsilon} becomes $H^{({\rm H})} _{mn{\bf k}} = \left<
u^{({\rm H})}_{m{\bf k}}\right| H_{\bf k} \left| u^{({\rm H})}_{n{\bf
    k}}\right> = \varepsilon^{({\rm H})}_{m{\bf k}}\, \delta_{mn}$.
For an isolated group of bands, or within the inner energy window in
the non-isolated case, the states given by Eq.~\eqref{eq:U} are
virtually identical to the corresponding eigenstates of the full
Hamiltonian $H_{\bf k}$. Hence they satisfy
\begin{equation}
H_{\bf k}\left| u^{({\rm H})}_{n{\bf k}}\right> \doteq
\varepsilon^{({\rm H})}_{n{\bf k}}
\left| u^{({\rm H})}_{n{\bf k}}\right>\,,
\label{eq:Hu-H}
\end{equation}
where the symbol
$\doteq$ denotes an equality that is strictly valid only for an
isolated set of bands or, when disentanglement is performed, within
the inner energy window. We shall make frequent use of the relation
\begin{equation}
H_{\bf k}\left| u^{({\rm W})}_{n\bf k}\right>\doteq
\sum_l\,\left| u^{({\rm W})}_{l\bf k}\right>H^{({\rm W})}_{ln\bf k}\,,
\label{eq:Hu-W}
\end{equation}
which follows from Eqs.~\eqref{eq:epsilon}--\eqref{eq:Hu-H},
\begin{eqnarray}
H_{\bf k}\left| u^{({\rm W})}_{n\bf k}\right>&=&
H_{\bf k}\sum_p\,\left| u^{({\rm H})}_{p\bf k}\right>
\left(U^\dagger_{\bf k}\right)_{pn}\nonumber\\
&\doteq&\sum_p\,
\left| u^{({\rm H})}_{p\bf k}\right>
\varepsilon^{({\rm H})}_{p\bf k}
\left(U^\dagger_{\bf k}\right)_{pn}\nonumber\\               
&=&\sum_l\,\left| u^{({\rm W})}_{l\bf k}\right>
\left[\sum_p\,
U_{lp\bf k}\varepsilon^{({\rm H})}_{p\bf k}\left(U^\dagger_{\bf k}\right)_{pn}
\right]\nonumber\\
&=&\sum_l\,\left| u^{({\rm W})}_{l\bf k}\right>H^{({\rm W})}_{ln\bf k}\,.
\end{eqnarray}

\subsubsection{Velocity matrix elements}
\label{sec:vel}

The same interpolation strategy can be applied to the velocity matrix
elements
$v_{y,mn{\bf k}}=\left< u_{m{\bf k}}\right| v_{y,{\bf k}} \left|
u_{n{\bf k}} \right>$ appearing in Eq.~\eqref{eq:sigma}. The
interpolated matrix elements are obtained with the help of
Eq.~\eqref{eq:U},
\begin{equation}
v_{y,mn{\bf k}}^{({\rm H})}=
\left< 
  u^{({\rm H})}_{m{\bf k}}\right| v_{y,{\bf k}} \left| u^{({\rm H})}_{n{\bf k}}
\right>
=
\left[
  U^\dagger_{\bf k}\, v^{({\rm W})}_{y,{\bf k}}\, U_{\bf k}
\right]_{mn}\,,
\label{eq:vyH}
\end{equation}
where
\begin{equation}
\hbar v^{({\rm W})}_{y,mn{\bf k}}=
\left<
  u^{({\rm W})}_{m{\bf k}}\right| \left(\partial_y H_{\bf k}\right)
  \left| u^{({\rm W})}_{n{\bf k}}
\right>.
\label{eq:vyW-a}
\end{equation}
We now expand the right-hand-side as
\begin{widetext}
\begin{equation}
\hbar v^{({\rm W})}_{y,mn{\bf k}}=H^{{\rm(W)}} _{y,mn{\bf k}}-
\left< \partial_y u^{({\rm W})}_{m{\bf k}} \right| H_{\bf k}
\left|u^{({\rm W})}_{n{\bf k}} \right>-
\left< u^{({\rm W})}_{m{\bf k}} \right| H_{\bf k}
\left|\partial_y u^{({\rm W})}_{n{\bf k}} \right>
\label{eq:vyW-b}
\end{equation}
\end{widetext}
where
$H^{{\rm(W)}}_{y,mn{\bf k}}=\partial_y H^{{\rm(W)}}_{mn{\bf k}}$, and
then use Eq.~\eqref{eq:Hu-W} to write
\begin{equation}
\left< \partial_y u^{({\rm W})}_{m{\bf k}} \right| H_{\bf k}
\left|u^{({\rm W})}_{n{\bf k}} \right>\doteq
i\sum_l\,A^{({\rm W})}_{y,ml{\bf k}}H^{({\rm W})}_{ln{\bf k}}\,,
\label{eq:HyW}
\end{equation}
where
\begin{equation}
A^{({\rm W})}_{y,mn{\bf k}}=i
\left.\left<u^{({\rm W})}_{m{\bf k}}
\right|\partial_y u^{({\rm W})}_{n{\bf k}} \right>
=\left[ A^{({\rm W})}_{y,nm{\bf k}} \right]^*
\label{eq:AWymn1}
\end{equation}
is the Berry connection matrix.  Combining Eqs.~\eqref{eq:vyH},
\eqref{eq:vyW-b}, and \eqref{eq:HyW} and introducing the notation
\begin{equation}
\overline{O}^{({\rm H})}_{\bf k}=U^\dagger_{\bf k}O^{({\rm W})}_{\bf k}U_{\bf k}\,,
\label{eq:O-bar}
\end{equation}
we arrive at the desired expression for the interpolated velocity
matrix elements [see also Eq.~(31) in Ref.~\onlinecite{Wang2006PRB}],
\begin{equation}
v_{y,mn{\bf k}}^{({\rm H})}\doteq\frac{1}{\hbar}\overline{H}^{({\rm H})}_{y,mn{\bf k}}-
\frac{i}{\hbar}
\left( \varepsilon^{({\rm H})}_{n{\bf k}} - \varepsilon^{({\rm H})}_{m{\bf k}} \right)
\overline{A}^{({\rm H})}_{y,mn{\bf k}}\,.
\label{eq:vyH-final}
\end{equation}
To evaluate it we use
\begin{equation}
  H^{{\rm(W)}} _{y,mn{\bf k}}=
  i\sum_{\bf R}\,e^{i{\bf k}\cdot {\bf R}} R_y
  \left<{\bf 0}m\right| H \left|{\bf R}n\right>
\label{eq:yHmn}
\end{equation}
which follows from differentiating Eq.~\eqref{eq:Hmn}, together with
\begin{equation}
  A^{{\rm(W)}} _{y,mn{\bf k}}=\sum_{\bf R}\,e^{i{\bf k}\cdot {\bf R}}
  \left<{\bf 0}m\right| y \left|{\bf R}n\right>
\label{eq:AWymn2}
\end{equation}
which is obtained by inserting Eq.~\eqref{eq:U} in
Eq.~\eqref{eq:AWymn1}.  These two matrices are then transformed
according to Eq.~\eqref{eq:O-bar}, using the matrix $U_{\bf k}$ from
Eq.~\eqref{eq:epsilon}.  As a reminder, the symbol $\doteq$ in
Eq.~\eqref{eq:vyH-final} means that when the MLWFs are generated from
a non-isolated group of bands, the interpolated velocity matrix
elements are only accurate when both
$\varepsilon^{({\rm H})}_{n{\bf k}}$ and
$\varepsilon^{({\rm H})}_{m{\bf k}}$ fall within the inner energy
window.

\subsubsection{Spin-current matrix elements}
\label{sec:spin-current}

The spin-current matrix elements
$j_{x,mn{\bf k}}^z= \left <u_{m\bf k}\right|j_x^z\left| u_{n\bf
  k}\right>$ appearing in Eq.~\eqref{eq:sigma} are interpolated as
\begin{equation}
j^{z\,({\rm H})}_{x,mn{\bf k}} =
\left[
  U^\dagger_{\bf k}\, j^{z\,({\rm W})}_{x,{\bf k}}\, U_{\bf k}
\right]_{mn}\,,
\label{eq:jzxmn_H}
\end{equation}
where, according to Eq.~\eqref{eq:jzx},
\begin{eqnarray}
j^{z\,({\rm W})} _{x,mn{\bf k}}&=&\frac{1}{2}
\left[\left< u^{({\rm W})}_{m{\bf k}}\right| s^z v_{x,{\bf k}}
\left| u^{({\rm W})}_{n{\bf k}}\right>\right.\nonumber\\
&+&
\left.\left< u^{({\rm W})}_{n{\bf k}}\right| s^z v_{x,{\bf k}}
\left| u^{({\rm W})}_{m{\bf k}}\right>^*\right]\,.
\label{eq:jzxmn}
\end{eqnarray}
For the ensuing manipulations it will be convenient to define the spin-related matrices
\begin{equation}
s^{z\,{\rm(W)}} _{mn{\bf k}}=
\left< u^{({\rm W})}_{m{\bf k}}\right| s^z \left| u^{({\rm W})}_{n{\bf k}}\right>
\label{eq:szW}
\end{equation}
and
\begin{equation}
{\mathcal S}^{z\,{\rm(W)}} _{x,mn{\bf k}}=
i\left< u^{({\rm W})}_{m{\bf k}}\right| s^z
\left| \partial_x u^{({\rm W})}_{n{\bf k}}\right>
\label{eq:SzcW}
\end{equation}
[the latter
is the spin version of the
Berry connection matrix in Eq.~\eqref{eq:AWymn1}], and to
note that they satisfy
\begin{align}
&\bigmatel{ \partial_x u^{({\rm W})}_{m{\bf k}} }{ s^z }{ u^{({\rm W})}_{n{\bf k}} } \nonumber\\
&=
\partial_x \bigmatel{ u^{({\rm W})}_{m{\bf k}} }{ s^z }{ u^{({\rm W})}_{n{\bf k}} }
- \bigmatel{ u^{({\rm W})}_{m{\bf k}} }{ s^z }{ \partial_x u^{({\rm W})}_{n{\bf k}} }\nonumber\\
&= \partial_x s^{z\,{\rm(W)}} _{mn{\bf k}} + i{\mathcal S}^{z\,{\rm(W)}} _{x,mn{\bf k}}\,.
\label{eq:SzcW2}
\end{align}
In analogy with Eq.~\eqref{eq:vyW-b}, we write the matrix
elements appearing in Eq.~\eqref{eq:jzxmn} as
\begin{eqnarray}
\hbar\left< u^{({\rm W})}_{m{\bf k}}\right| s^z v_{x,{\bf k}}
\left| u^{({\rm W})}_{n{\bf k}}\right>&=&
\partial_x \left< u^{({\rm W})}_{m{\bf k}}\right| s^z H_{\bf k}
\left| u^{({\rm W})}_{n{\bf k}}\right>\nonumber\\
&-&\left< \partial_x u^{({\rm W})}_{m{\bf k}}\right| s^z H_{\bf k}
\left| u^{({\rm W})}_{n{\bf k}}\right>\nonumber\\
&-&\left< u^{({\rm W})}_{m{\bf k}}\right| s^z H_{\bf k} \left|
\partial_x u^{({\rm W})}_{n{\bf k}}\right>\,,
\end{eqnarray}
and then use Eqs.~\eqref{eq:Hu-W} and~\eqref{eq:SzcW2}
to arrive at
\begin{widetext}
\begin{equation}
\hbar\left< u^{({\rm W})}_{m{\bf k}}\right| s^z v_{x,{\bf k}}
\left| u^{({\rm W})}_{n{\bf k}}\right>\doteq
\left[
s^{z\,{\rm(W)}} _{\bf k}
H^{{\rm(W)}} _{x,{\bf k}}
-
i{\mathcal S}^{z\,{\rm(W)}} _{x,{\bf k}}H^{{\rm(W)}} _{\bf k}
\right]_{mn}
-\left< u^{({\rm W})}_{m{\bf k}}\right| s^z H_{\bf k}
\left| \partial_x u^{({\rm W})}_{n{\bf k}}\right>\,.
\label{eq:svW}
\end{equation}
\end{widetext}
The interpolated spin-current matrix elements are obtained by
combining Eqs.~\eqref{eq:jzxmn_H}, \eqref{eq:jzxmn}, and
\eqref{eq:svW}.  In addition to Eq.~\eqref{eq:Hmn} for
$H_{\bf k}^{({\rm W})}$ and Eq.~\eqref{eq:yHmn}
for $H_{x,{\bf k}}^{({\rm W})}$, the following
Fourier sums are needed to evaluate Eq.~\eqref{eq:svW},

\begin{equation}
s^{z\,{\rm(W)}} _{mn{\bf k}}=\sum_{\bf R}\,e^{i{\bf k}\cdot {\bf R}}
\left<{\bf 0}m\right|s^z\left|{\bf R}n\right>\,,
\label{eq:szmn}
\end{equation}
\begin{equation}
{\mathcal S}^{z\,{\rm(W)}} _{x,mn{\bf k}}=
\sum_{\bf R}\,e^{i{\bf k}\cdot {\bf R}}
\left<{\bf 0}m\right|s^z(x-R_x)\left|{\bf R}n\right>\,,
\label{eq:szxmn}
\end{equation}
and
\begin{eqnarray}
&&\left< u^{({\rm W})}_{m{\bf k}}\right| s^z H_{\bf k}
\left| \partial_x u^{({\rm W})}_{n{\bf k}}\right>\nonumber\\
&=&-i\sum_{\bf R}\,e^{i{\bf k}\cdot {\bf R}}
\left<{\bf 0}m\right|s^z H (x-R_x)\left|{\bf R}n\right>\,.
\label{eq:szHxmn}
\end{eqnarray}

\subsection{Real-space matrix elements}
\label{sec:real-space}

The calculations described above require the following real-space
matrix elements,
\begin{eqnarray}
\left<{\bf 0}m\right| H \left|{\bf R}n\right>\,,
\left<{\bf 0}m\right| y \left|{\bf R}n\right>\,,
\left<{\bf 0}m\right|s^z\left|{\bf R}n\right>\,,\nonumber\\
\left<{\bf 0}m\right|s^z  (x-R_x)\left|{\bf R}n\right>\,,
\left<{\bf 0}m\right|s^z H (x-R_x)\left|{\bf R}n\right>\,.\nonumber\\
\label{eq:r-list}
\end{eqnarray}
The first one is needed to interpolate the energy eigenvalues, the
first two are needed for the velocity matrix elements, and all but the
second are needed for the spin-current matrix elements. In practice,
we evaluate them as inverse Fourier transforms over the {\it ab
  initio} BZ grid $\{{\bf q}\}$ that was used when constructing the
MLWFs.

The output of the wannierization procedure is a set of unitary or
semi-unitary matrices (for isolated or non-isolated groups of bands,
respectively) relating the Bloch-sum states
$\ket{u^{({\rm W})}_{n{\bf q}}}$
to the {\it ab initio}
eigenstates
$\left|u_{n{\bf
    q}}\right>$.\cite{Marzari1997PRB,Souza2001PRB,mostofi2008,Marzari2012RMP}
Together with the {\it ab initio} energy eigenvalues, those matrices
can be used to build the matrices $H^{({\rm W})}_{\bf q}$, from which
the real-space matrix Hamiltonian elements can then be obtained by
inverting Eq.~\eqref{eq:Hmn},
\begin{equation}
\left<{\bf 0}m\right| H \left|{\bf R}n\right>=\frac{1}{N_q}
\sum_{\bf q} e^{-i{\bf q}\cdot {\bf R}}H^{({\rm W})}_{mn{\bf q}}
\label{eq:0HR}
\end{equation}
(here, $N_q$ is the number of ${\bf q}$ points).  Similarly, inverting
Eq.~\eqref{eq:szmn} gives
\begin{equation}
\left<{\bf 0}m\right|s^z\left|{\bf R}n\right>= \frac{1}{N_q}
\sum_{\bf q} e^{-i{\bf q}\cdot {\bf R}}s^{z\,{\rm(W)}}_{mn{\bf q}}\,,
\label{eq:0szR}
\end{equation}
where once again the spin matrix elements are first evaluated between
{\it ab initio} states, and then converted to matrix elements between
Bloch-sum states.

We now turn to the matrix elements containing the coordinate operators
$x$ and $y$. Inverting Eq.~\eqref{eq:AWymn2} we get
\begin{equation}
\left<{\bf 0}m\right| y \left|{\bf R}n\right> =
\frac{1}{N_q}\sum_{\bf q}\,e^{-i{\bf q}\cdot {\bf R}} A^{{\rm(W)}}_{y,mn{\bf q}}
\label{eq:0yR}
\end{equation}
where, from Eq.~\eqref{eq:AWymn1},
\begin{equation}
A^{({\rm W})}_{y,mn{\bf q}}=
i\left.\left<u^{({\rm W})}_{m{\bf q}} \right|
\partial_y u^{({\rm W})}_{n{\bf q}} \right>.
\label{eq:AWymnq}
\end{equation}
As discussed below Eq.~\eqref{eq:BlochSum}, the cell-periodic
Bloch-sum states are smooth functions of the crystal momentum. This
allows us to use a finite-difference representation on the
$\{{\bf q}\}$ grid of the differential operator in
Eq.~\eqref{eq:AWymnq}.  Following Appendix~B of
Ref.~\onlinecite{Marzari1997PRB} we write
\begin{equation}
\left| \partial_{\bf q} u^{({\rm W})}_{n{\bf q}} \right>\approx
\sum_{\bf b}\,w_b{\bf b}
\left[
\left| u^{({\rm W})}_{n,{\bf q}+{\bf b}} \right>-
\left| u^{({\rm W})}_{n{\bf q}} \right>
\right]\,,
\label{eq:finite-diff}
\end{equation}
where the vector {\bf b} connects $\mathbf{q}$ to its nearest-neighbor grid points, and
$w_b$ is an appropriate weight factor that only depends on
$b=|{\bf b}|$. Inserting Eq.~\eqref{eq:finite-diff} in
Eq.~\eqref{eq:AWymnq} yields
\begin{eqnarray}
A^{({\rm W})}_{y,mn}({\bf q})&\approx&
i\sum_{\bf b}w_b b_y \left(\left.\left<u^{({\rm W})}_{m{\bf q}} \right|
u^{({\rm W})}_{n,{\bf q}+{\bf b}} \right>-\delta_{mn}\right)\nonumber\\
&=&i\sum_{\bf b}w_b b_y \left.\left<u^{({\rm W})}_{m{\bf q}} \right|
u^{({\rm W})}_{n,{\bf q}+{\bf b}} \right>\,.
\end{eqnarray}

The remaining real-space matrix elements are evaluated in a similar
manner, by inverting Eqs.~\eqref{eq:szxmn} and~\eqref{eq:szHxmn} and
then using Eq.~\eqref{eq:finite-diff}. This leads to
\begin{equation}
\left<{\bf 0}m\right|s^z(x-R_x)\left|{\bf R}n\right>=
\frac{1}{N_q}\sum_{\bf R}\,e^{-i{\bf q}\cdot {\bf R}}
{\mathcal S}^{z\,{\rm(W)}} _{x,mn{\bf k}}
\label{eq:0SzyR}
\end{equation}
where
\begin{eqnarray}
{\mathcal S}^{z\,{\rm(W)}} _{x,mn{\bf k}}\approx
i\sum_{\bf b}w_b b_y
\left<u^{({\rm W})}_{m{\bf q}} \right|s^z
\left|u^{({\rm W})}_{n,{\bf q}+{\bf b}} \right>\,,
\label{eq:szvx}
\end{eqnarray}
and
\begin{eqnarray}
&&\left<{\bf 0}m\right|s^z H (x-R_x)\left|{\bf R}n\right>
\nonumber\\
&=&\frac{i}{N_q}\sum_{\bf R}\,e^{-i{\bf q}\cdot {\bf R}}
\left< u^{({\rm W})}_{m{\bf q}}\right| s^z H_{\bf q}
\left| \partial_x u^{({\rm W})}_{n{\bf q}}\right>
\label{eq:0szHyR}
\end{eqnarray}
where
\begin{eqnarray}
&&\left< u^{({\rm W})}_{m{\bf q}}\right| s^z H_{\bf q}
\left| \partial_x u^{({\rm W})}_{n{\bf q}}\right> \nonumber\\
&\approx&
\sum_{\bf b}w_b b_x
\left<u^{({\rm W})}_{m{\bf q}} \right|s^z H_{\bf q}
\left|u^{({\rm W})}_{n,{\bf q}+{\bf b}} \right>\,.
    \label{eq:szHvx}
\end{eqnarray}

The current release of {\tt Wannier90} already provides the first
three real-space matrix elements in
Eq.~\eqref{eq:r-list},~\cite{mostofi2014updated}
and for this work we have implemented the other two.
As before, we first obtain the relevant $k$-space matrix elements between {\it
  ab~initio} eigenstates at neighboring grid points, then perform
a (semi)-unitary transformation to find the corresponding matrix
elements between Bloch-sum states, and finally we Fourier-transform
these to real space via Eqs.~\eqref{eq:0SzyR}
and~\eqref{eq:0szHyR}.

This concludes the description of the Wannier-based interpolation
scheme for evaluating the intrinsic SHC. To summarize, at
each point ${\bf k}$ in Eq.~\eqref{eq:sigma} one replaces the energy
eigenvalues, velocity, and spin-current matrix elements by their
interpolated counterparts, letting the summations over band indices
run from $1$ to $N_{\rm W}$. Strictly speaking those summations should
be further restricted to states within the inner energy window, but
not doing so introduces a negligible error: by virtue of the energy
denominator squared in Eq.~\eqref{eq:sigma}, the SHC is strongly
dominated by contributions from pairs of nearby occupied and empty
states within the inner window.

\subsection{Approximations to the Wannier-interpolation scheme}
\label{sec:approx}
In this section we discuss two approximations that can be used to
simplify the evaluation of the intrinsic SHC by Wannier interpolation,
as indeed has been done in some previous works.\cite{Feng2012PRB,Sui2017PRB,
Sun2016PRL,Zhang2017PRB,Zelezny2017PRL,derunova2019sciadv,Zhang2018NJP}

\subsubsection{Projected-spin approximation}
\label{sec:projected-spin}

The first approximation only affects the spin-current
matrix elements, and it amounts to replacing Eq.~\eqref{eq:svW} by
\begin{equation}
\left< u^{({\rm W})}_{m{\bf k}}\right| s^z v_{x,{\bf k}}
\left| u^{({\rm W})}_{n{\bf k}}\right>\approx
\left[
s^{z\,{\rm(W)}}_{\bf k} v^{({\rm W})}_{x,{\bf k}}
\right]_{mn}\,,
\label{eq:svW_approx}
\end{equation}
where the matrices $v^{({\rm W})}_{x,{\bf k}}$ and
$s^{z\,{\rm(W)}}_{\bf k}$ are defined by Eqs.~\eqref{eq:vyW-a}
and~\eqref{eq:szW}, respectively.  This approximation is valid
provided that the state $s^z\left| u^{({\rm W})}_{m\bf k}\right>$
remains within the projected Wannier subspace at ${\bf k}$, and
henceforth we will call it the ``projected-spin approximation.''

Equation~\eqref{eq:svW_approx} has the practical advantage over
Eq.~\eqref{eq:svW} that it allows to evaluate the SHC using only the
first three matrix elements in Eq.~\eqref{eq:r-list}, which are
readily available in the current release of {\tt Wannier90}.

In the following, we analyze the validity of the projected-spin
approximation for calculating the intrinsic SHC. We do so under the assumption that the spin Hall effect is mediated by
the spin-orbit interaction. This is true for the nonmagnetic systems
considered in this work, as well as for systems with collinear magnetic
order. It is not true, however, for systems with noncollinear spin textures, where the spin Hall effect can occur regardless of the spin-orbit coupling strength.\cite{Zhang2018NJP}

Under the above assumption,
the projected-spin approximation should be valid provided that the
spin-orbit interaction does not mix two electronic states when one
lies inside the Wannier manifold, and the other outside. To see this,
let $\left|u^{({\rm W})}_{(s,n_s){\bf k}}\right>$ be a spin-polarized
Bloch-sum state calculated without spin-orbit coupling
for some choice of the spin quantization axis.
Here $s=\pm 1$ is the spin index,
and $n_s$ is the band
index for a given spin. Now
turn on the spin-orbit interaction,
and write the new Bloch-sum states as
$\left|u^{({\rm W})}_{m{\bf k}}\right>$.
If the spin-orbit interaction is sufficiently
weak that it only mixes states within the Wannier manifold,
we have
\begin{equation}
\left|u^{({\rm W})}_{m{\bf k}}\right>\approx \sum_{s,n_s}
\left|u^{({\rm W})}_{(s,n_s){\bf k}}\right>\,\widetilde{U}_{(s,n_s)\, m{\bf k}}
    \label{eq:Utilde_small_soc}
\end{equation}
with an appropriate choice of the unitary matrix $\widetilde{U}_{\bf k}$. 
From this relation and its inverse
we obtain
\begin{eqnarray}
&&s^z \left|u^{({\rm W})}_{m{\bf k}}\right>\approx \sum_{s,n_s}\,
s\left|u^{({\rm W})}_{(s,n_s){\bf k}}\right>\,
\widetilde{U}_{(s,n_s)m{\bf k}}
\label{eq:szu_small_soc}\nonumber\\
&\approx&\sum_l \left|u^{({\rm W})}_{l{\bf k}}\right>
\left[
\sum_{s,n_s} \left(\widetilde{U}_{\bf k}^\dagger\right)_{l(s,n_s)}\,s\,
\widetilde{U}_{(s,n_s)m{\bf k}}
\right] \nonumber\\
&=& \sum_l \left|u^{({\rm W})}_{l{\bf k}}\right>
\left[s_\mathbf{k}^{z(\textrm{W})}\right]_{lm}\,,
\label{eq:sz_is_unitary}
\end{eqnarray}
which leads the right-hand-side of Eq.~\eqref{eq:svW_approx} when
inserted on the left-hand side.

Even if spin-orbit coupling is not weak, the mixing via the spin
operator between states inside and outside the Wannier manifold only
occurs near the energy bounds of that manifold.  But since the
calculated SHC should be converged with respect to the number
$N_{\rm W}$ of wannierized bands, the only practical disadvantage of
the projected-spin approximation is
a slower convergence of the SHC with respect to $N_{\rm W}$.

\subsubsection{Tight-binding approximation}
\label{sec:diagonal-r}

The SHC of a tight-binding model in which every basis orbital
is considered to be perfectly localized at a point in 
space can be obtained by assuming that
\begin{equation}
\mathbf{r} \left|{\bf R}n\right> \approx
(\mathbf{R}+\boldsymbol{\tau}_n) \left|{\bf R}n\right> \,,
\label{eq:0yR_TB}
\end{equation}
where ${\boldsymbol \tau}_n=\langle {\bf 0}n\vert {\bf r}\vert {\bf 0}n\rangle$. We will call this the ``tight-binding approximation.'' It leads to [see Eq.~\eqref{eq:AWymn2}]
\begin{equation}
{\bf A}^{{\rm(W)}} _{mn{\bf k}}\approx {\boldsymbol\tau}_{m}\,\delta_{mn}\,,
\label{eq:AWymn2_TB}
\end{equation}
and finally [see Eqs.~\eqref{eq:vyW-b} and~\eqref{eq:HyW}] to
\begin{equation}
\hbar{\bf v}^{({\rm W})}_{mn{\bf k}}\approx
{\boldsymbol\nabla}_{\bf k}H^{{\rm(W)}} _{mn{\bf k}}
-i\left({\boldsymbol\tau}_m-{\boldsymbol\tau}_n\right)\,
H^{{\rm(W)}} _{mn{\bf k}}\,.
\label{eq:vW_TB}
\end{equation}
This is the standard expression for the velocity
matrix elements in the empirical tight-binding method.  Since a
tight-binding model only has on-site energies and inter-site hopping
integrals, it cannot reproduce, for example, the velocity matrix
element between two Bloch-sum states when one is made up of atomic-like $s$ orbitals and the other of atomic-like $p$ orbitals.
The tight-binding approximation is best suited for crystals with
larger energy band widths, in which the dominant contribution to
Eq.~\eqref{eq:vW_TB} comes from inter-site hopping integrals.

Interestingly, the tight-binding approximation~\eqref{eq:0yR_TB}
implies the projected-spin approximation~\eqref{eq:svW_approx}, so that it is not
meaningful to assume the former without assuming the latter.
To show this, we first note that in the tight-binding approximation
the following equality holds,
\begin{equation}
\boldsymbol{\nabla}_{\bf k} \ket{u_{n\mathbf{k}}^{(\textrm{W})}}
\approx
-i \boldsymbol{\tau}_n \ket{u_{n\mathbf{k}}^{(\textrm{W})}} \,,
\label{eq:dunk_TB}
\end{equation}
as can be seen by differentiating Eq.~\eqref{eq:BlochSum} with respect to~${\bf k}$ and then invoking Eq.~\eqref{eq:0yR_TB}.
We now use this relation together with  Eq.~\eqref{eq:Hu-W} in the identity
\begin{equation}
\hbar\mathbf{v}_{\bf k} \ket{u_{n\mathbf{k}}^{(\textrm{W})}}
= \boldsymbol{\nabla}_{\bf k} 
\left[ H_{\bf k} \ket{u_{n\mathbf{k}}^{(\textrm{W})}} \right]
- H_{\bf k} \boldsymbol{\nabla}_{\bf k} \ket{u_{n\mathbf{k}}^{(\textrm{W})}}
\end{equation}
to obtain
\begin{equation}
\hbar {\bf v}_{\bf k}\ket{u_{n\mathbf{k}}^{(\textrm{W})}}\dotapprox
\sum_m\,\ket{u_{m\mathbf{k}}^{(\textrm{W})}}
\left[
    \boldsymbol{\nabla}_{\bf k}H^{(\textrm{W})}_{mn{\bf k}}-
    i(\boldsymbol{\tau}_m-\boldsymbol{\tau}_n)H^{(\textrm{W})}_{mn{\bf k}}
\right]\,.
\end{equation}
Comparing with Eq.~\eqref{eq:vW_TB} we find that
\begin{equation}
\mathbf{v}_{\bf k}\ket{u_{n\mathbf{k}}^{(\textrm{W})}}
\dotapprox \sum_m\,\ket{u_{m\mathbf{k}}^{(\textrm{W})}} \mathbf{v}^{(\textrm{W})}_{mn\mathbf{k}}
\end{equation}
in the tight-binding approximation.
Inserting this relation on the left-hand-side
of Eq.~\eqref{eq:svW_approx} yields its right-hand-side, which
concludes the proof.

In practice,  the tight-binding approximation to the
calculation of the SHC by Wannier interpolation amounts to 
replacing Eq.~\eqref{eq:AWymn2} with Eq.~\eqref{eq:AWymn2_TB}
({\it i.e.}, discarding
 off-diagonal matrix elements of the coordinate operator in the
Wannier basis), and Eq.~\eqref{eq:svW} with
Eq.~\eqref{eq:svW_approx}. It therefore affects both types of matrix
elements (velocity and  spin-current)
appearing in the Kubo formula of Eq.~\eqref{eq:sigma}.

\section{Computational Details}
\label{sec:comp-details}

The fully-relativistic electronic structure of fcc Pt and zincblende
GaAs was calculated by density-functional theory, using the plane-wave
pseudopotential code {\tt pwscf} from the {\tt Quantum Espresso}
package.\cite{giannozzi2009quantum} The exchange-correlation energy
was approximated using the {\tt PBEsol}
functional.\cite{PerdewPRL2008}
Norm-conserving pseudopotentials were used for all atoms,
and the energy cutoff for the
wavefunctions was set to 60 Ry in the case of Pt and to 80 Ry in the
case of GaAs.
 We first performed self-consistent ground-state
calculations with the experimental lattice parameters
where the BZ was sampled on a uniform $12 \times 12 \times 12$
grid.
Non-self-consistent calculations were then carried out on
$8 \times 8 \times 8$ grids to obtain the Bloch eigenstates from
which the MLWFs were constructed
using the {\tt Wannier90}
package.\cite{mostofi2008,mostofi2014updated}

For both materials, 18 spinor MLWFs per cell were obtained using a
two-step procedure: (i) subspace selection (disentanglement), and (ii)
gauge selection within the disentangled
subspace.\cite{Marzari1997PRB,Souza2001PRB,mostofi2008,Marzari2012RMP}
In the case of Pt we used
$s$, $p$, and $d$ atom-centered nodeless trial orbitals, and the inner energy
window for the disentanglement step was chosen to span an energy range
of 15~eV
 from the bottom of
the valence bands up to 4~eV above the Fermi level.
For GaAs we used $s$ and $p$ atom-centered orbitals, together with
$s$-like orbitals centered at tetrahedral interstitial sites;
the inclusion of the interstitial orbitals helped obtaining a
good description of the higher-energy part of the conduction bands.
The inner window went in this case from the bottom of the valence bands 
up to 7~eV above the valence-band maximum.  

The matrix elements between {\it ab initio} eigenstates that are needed to
obtain the real-space matrix elements in Eq.~\eqref{eq:r-list} were
calculated in {\tt pw2wannier90}, the interface between {\tt pwscf}
and {\tt Wannier90}. As mentioned earlier, some of those matrix elements were already available and the others were coded by us.

The Wannier-interpolation calculations of the static and dynamic SHC
were carried out on dense uniform $k$-point grids containing the
$\Gamma$ point. The calculations were repeated using different grids,
and the results were found to be well converged with
a $100 \times 100 \times 100$ grid in the case of Pt, and with a
$150 \times 150 \times 150$ grid in the case of GaAs.

When calculating the dynamic SHC, we employed the ``adaptive
smearing'' scheme of Ref.~\onlinecite{Yates2007PRB} in order to
capture the sharp features in the spectrum. In this scheme,
the parameter $\delta$ in Eq.~\eqref{eq:sigma}
is chosen as
$\delta=a\left| \partial E_{n\mathbf{k}}/\partial\mathbf{k} - \partial
E_{m\mathbf{k}}/\partial\mathbf{k}\right|\Delta k$, where $a$ is a
dimensionless constant of order one (we set $a=\sqrt{2}$),
and $\Delta k$ is the distance in $k$-space between nearest-neighbor
points on the interpolation grid.

\section{Results and Discussion}
\label{sec:results}

\subsection{
Energy bands of Pt and GaAs}
\label{sec:wannierization}

\begin{figure}
\includegraphics[width=0.88\columnwidth]{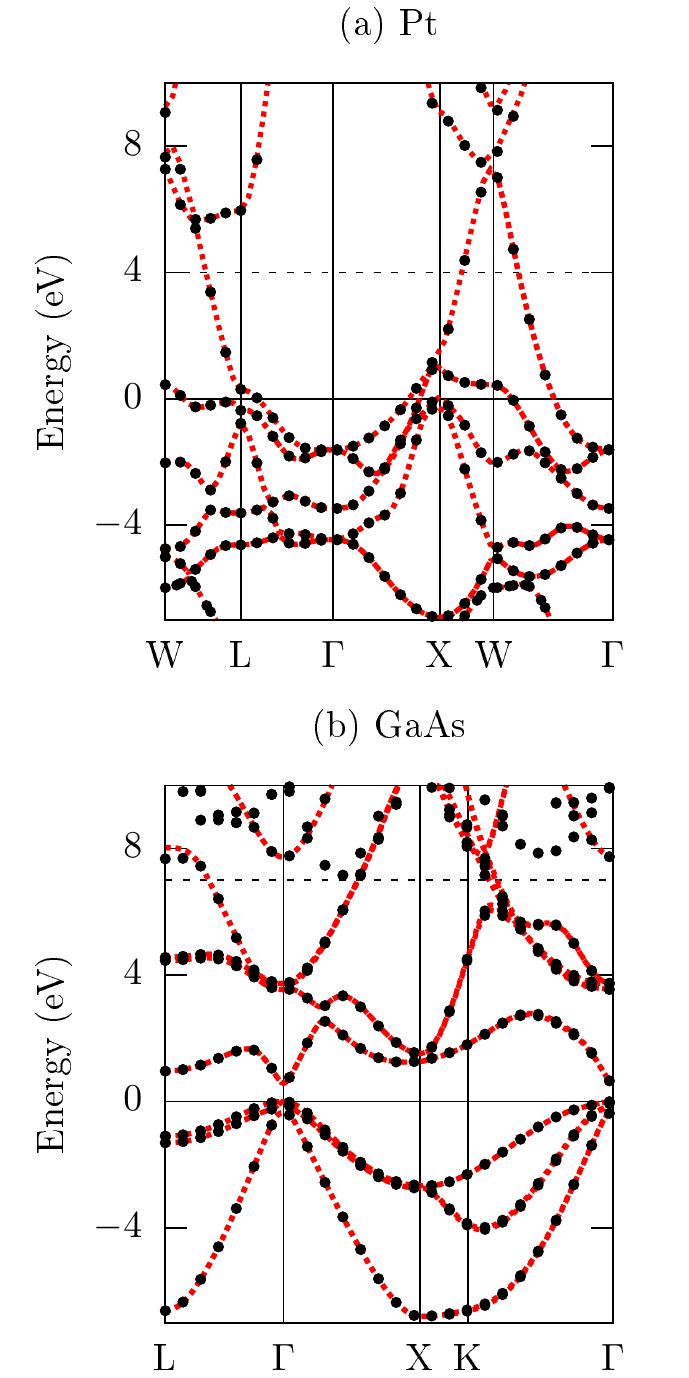}
\caption{The electronic band structures of (a) fcc Pt and (b)
  zincblende GaAs. The black dots are the energy eigenvalues obtained
  directly from \textit{ab initio} calculations, and the red dotted
  lines are the energy bands obtained by Wannier interpolation. The
  reference energy is indicated by a solid horizontal line.
    For Pt that reference energy is the Fermi level, and
  for GaAs it is the top of the valence band.
  In both cases, the inner window for the disentanglement step
  goes from the bottom of the valence bands up to the dashed horizontal line.
  }
\label{fig:bulkband}
\end{figure}

Figure~\ref{fig:bulkband} shows a comparision between the {\it ab
  initio} and Wannier-interpolated bandstructures of bulk Pt and
GaAs. It is clear that the chosen MLWFs faithfully reproduce the {\it
  ab initio} electronic states within a sizeable energy range around
the Fermi energy (Pt) or around the band gap (GaAs). Those MLWFs can
therefore be used to compute reliably the intrinsic SHC.

\begin{figure}
\includegraphics[width=0.9\columnwidth]{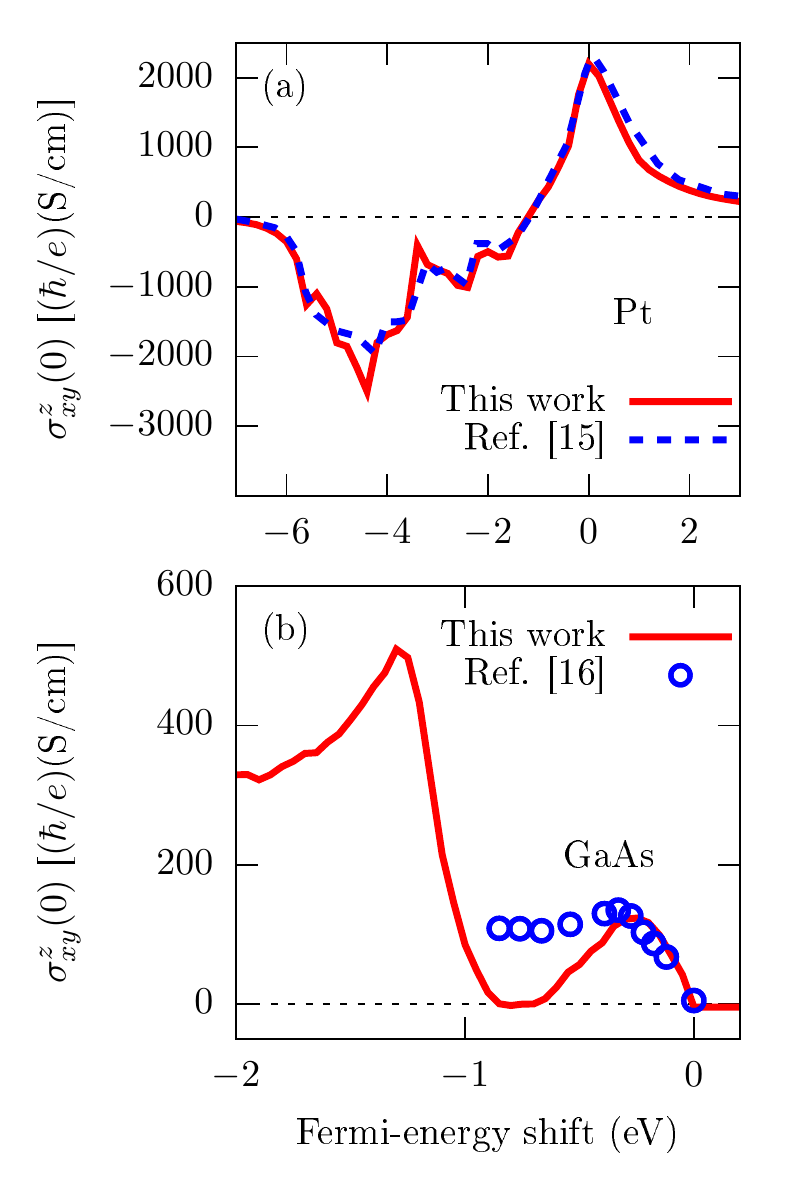}
\caption{Static SHC of (a) bulk Pt and (b) bulk GaAs calculated
    in this work as a function of the
  shift in the Fermi energy relative to its self-consistent
    value. The results of Refs.~\onlinecite{Guo2008PRL,Guo2005PRL}
  are also shown for comparison.
    }
\label{fig:shc_conv_comp}
\end{figure}
  
\subsection{Static spin Hall conductivities of Pt and GaAs}

\label{sec:shc_results}

The static SHC of Pt and GaAs is plotted in Fig.~\ref{fig:shc_conv_comp} as a function of the
shift in Fermi energy relative to its self-consistent value. 
Physically, this shift can be ascribed to a change in electron
density from either alloying (Pt) or doping (GaAs),
provided that the electronic structure does not change appreciably
in the process (the so-called ``rigid-band approximation'').

The results for Pt in Fig.~\ref{fig:shc_conv_comp}(a) are in good
agreement with those reported in a previous first-principles
study.\cite{Guo2008PRL} The large peaks around the unshifted Fermi level, and
$4$~eV below it, are caused by avoided band crossings induced by
the spin-orbit interaction.\cite{Guo2008PRL} For {\it p}-doped GaAs,
Fig.~\ref{fig:shc_conv_comp}(b) shows that the SHC calculated with our
method has a peak when the Fermi level is placed $0.3$~eV below the
top of the valence bands ($\Delta E_{\rm F}=-0.3~$eV), and that it
becomes nearly zero when $\Delta E_{\rm F}=-0.8$~eV. This is in
contrast to a previous first-princples study,\cite{Guo2005PRL} where
it was reported that the intrinsic SHC remains constant at around
$120 \,(\hbar/e)$~S/cm for $\Delta E_{\rm F}$ between $-0.3$~eV and
$-0.8$~eV, as also shown in the figure.

Let us now check the validity for Pt and GaAs of the
approximation schemes described in Sec.~\ref{sec:approx}, starting with
projected-spin approximation.
As discussed in Sec.~\ref{sec:projected-spin}, in the case of nonmagnetic systems
that approximation is valid  provided that the Bloch subspace spanned by the MLWFs is left invariant under the action
of the spin operator $s^z=(\hbar/2)\sigma^z$.
In the following, we quantity the extent to which this
condition is satisfied for each interpolated eigenstate
separetely.

We begin by resolving the identity operator as
\begin{equation}
\mathbbm{1}=
\sum_m
{\vert u^{(\mathrm{H})}_{m\mathbf{k}} \rangle \langle u^{(\mathrm{H})}_{m\mathbf{k}}
  \vert} + \sum_{r}{\vert v_{r\mathbf{k}} \rangle \langle v_{r\mathbf{k}}\vert}\,,
\label{eq:identity}
\end{equation}
where the states $\{\vert v_{r\mathbf{k}} \rangle \}$ span the space
of lattice-periodic functions orthogonal to every
$\ket{u^{(\mathrm{H})}_{m\mathbf{k}}}$ ($m=1,\ldots,N_{\rm W}$).
Inserting Eq.~\eqref{eq:identity} in the identity
$1= \left< u^{({\rm H})} _{n{\bf k}}\right| (\sigma^z)^2 \left|u^{({\rm
    H})} _{n{\bf k}}\right>$ we obtain
$1=I_{n\bf k}+R_{n\bf k}$,
where 
\begin{equation}
R_{n\bf k}=\sum_{r}\left|
\bigmatel{ v_{r{\bf k}} }{\sigma^z }{ u^{({\rm H})} _{n{\bf k}} }
\right|^2
\end{equation}
and
\begin{eqnarray}
I_{n{\bf k}}&\equiv&\sum_m
\left|\left< u^{({\rm H})}_{m{\bf k}}\right|\sigma^z
  \left|u^{({\rm H})}_{n{\bf k}}\right>\right|^2\nonumber\\
  &=&\sum_m \left|\left< u^{({\rm W})}_{m{\bf k}}\right|\sigma^z
  \left|u^{({\rm H})}_{n{\bf k}}\right>\right|^2
    \label{eq:Itilde}
\end{eqnarray}
are the weights of the normalized state
$\sigma^z\left| u^{({\rm H})}_{n\bf k}\right>$ inside and outside the
projected Wannier subspace, respectively.
Since those weights add up to one, any significant deviation of $I_{n{\bf k}}$
from unity
would indicate a failure in the projected-spin approximation.

\begin{figure}
\includegraphics[width=0.88\columnwidth]{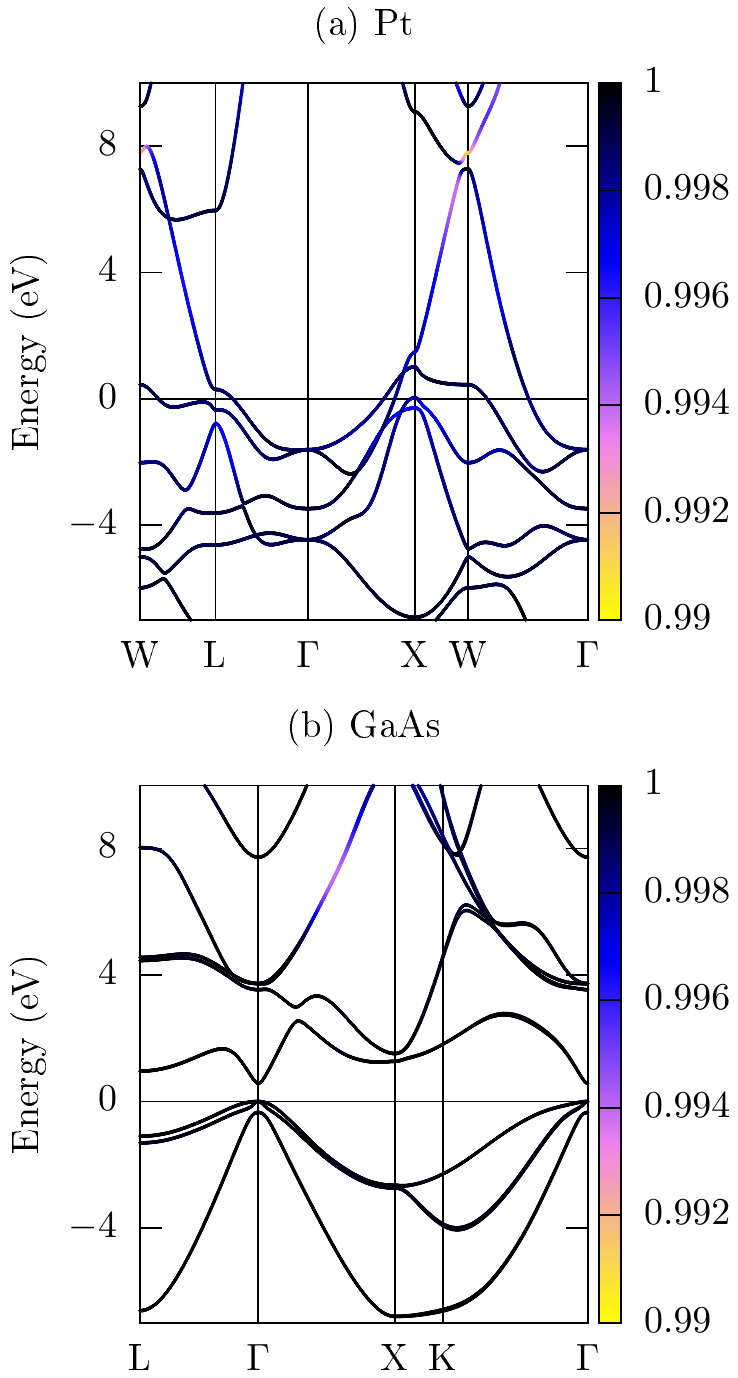}
\caption{ The interpolated band structures of (a) Pt and (b) GaAs, 
color-coded by the quantity
    $I_{n{\bf k}}$
    given by Eq.~\eqref{eq:Itilde}.}
\label{fig:sz_squared}
\end{figure}

\begin{figure*}
  \includegraphics[width=1.0\textwidth]{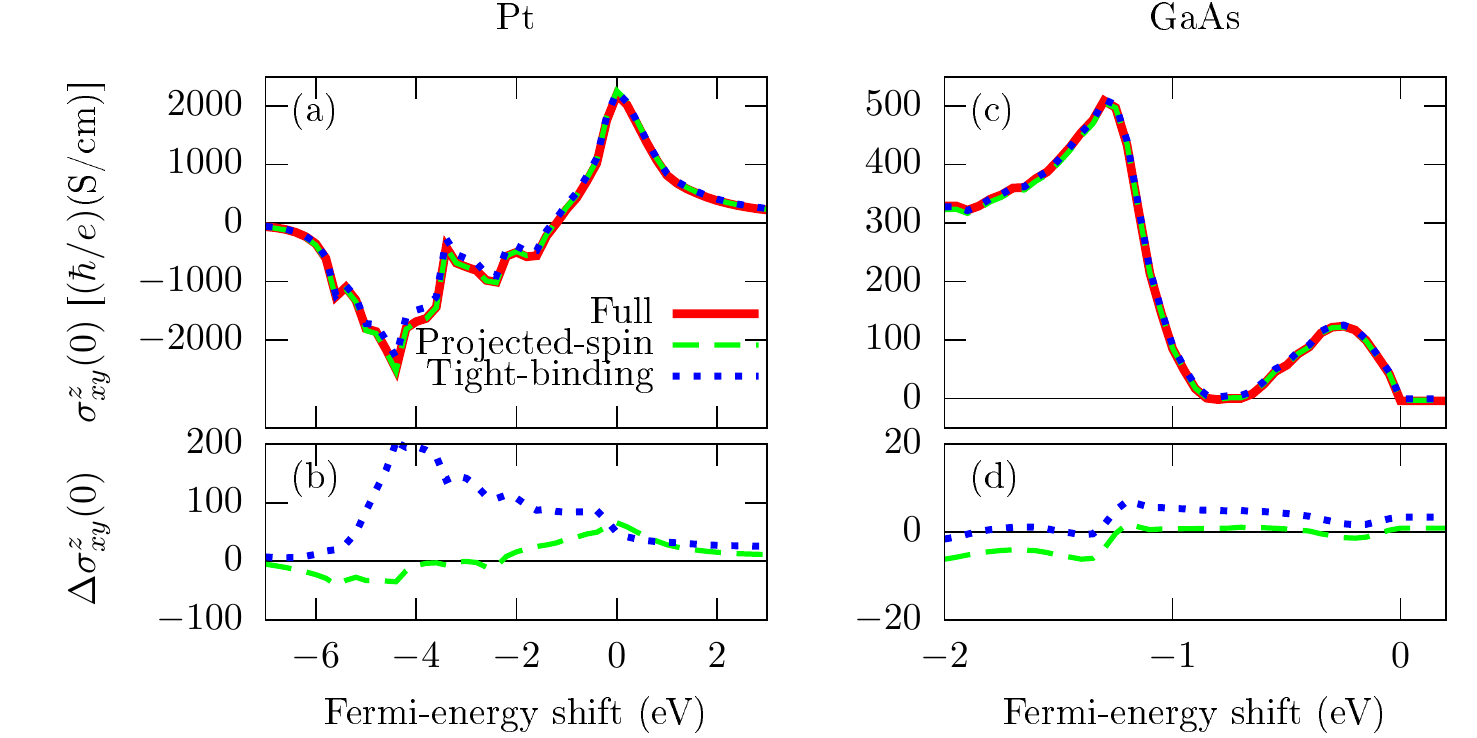}
  \caption{ (a) The calculated static SHC
    of bulk Pt as a function of the
    shift in the Fermi energy relative to its self-consistent
    value. The solid red  curve is the same as that in
    Fig.~\ref{fig:shc_conv_comp}(a), and it corresponds to a full Wannier-based
    calculation of the velocity and spin-current matrix elements, without any approximations.
    The dashed green curve corresponds
    to a calculation using the projected-spin
    approximation of Sec.~\ref{sec:projected-spin}, and
    the dotted blue curve corresponds
    to a calculation using the tight-binding approximation of
    Sec.~\ref{sec:diagonal-r}.
    (b) The difference in the SHC values between each of the two approximate
    calculations and the full one. (c) and (d) show the same
    quantities as (a) and (b), but for GaAs.
    }
  \label{fig:shc_approx}
\end{figure*}

Figure~\ref{fig:sz_squared} shows the interpolated energy bands of
Pt and GaAs color-coded by the quantity $I_{n{\bf k}}$; this quantity remains
very close to unity for every band that is well described by the
MLWFs,
suggesting that the projected-spin approximation
is valid for both materials.
This is confirmed by Fig.~\ref{fig:shc_approx}, where it can be seen
that the error that it introduces in the calculated SHC is of a few
percent at most.
(As discussed
earlier, that error can be systematically
reduced by increasing the number $N_{\rm W}$ of MLWFs per cell.)

Figure~\ref{fig:shc_approx} also shows the impact on the calculated SHC of the tight-binding approximation of Sec.~\ref{sec:diagonal-r}.
The error it introduces is again relatively small, of
the order of 10\%. 
The reliability of the tight-binding approximation in the context of the Wannier-interpolation scheme suggests that
the empirical tight-binding method, with parameters obtained by fitting
to either experimental results or
first-principles calculations, should yield
reasonably good results for the SHC in these classes of materials.

\subsection{Dynamic spin Hall conductivity of GaAs}
\label{sec:dyn_shc_results}
\begin{figure}
\includegraphics[width=0.9\columnwidth]{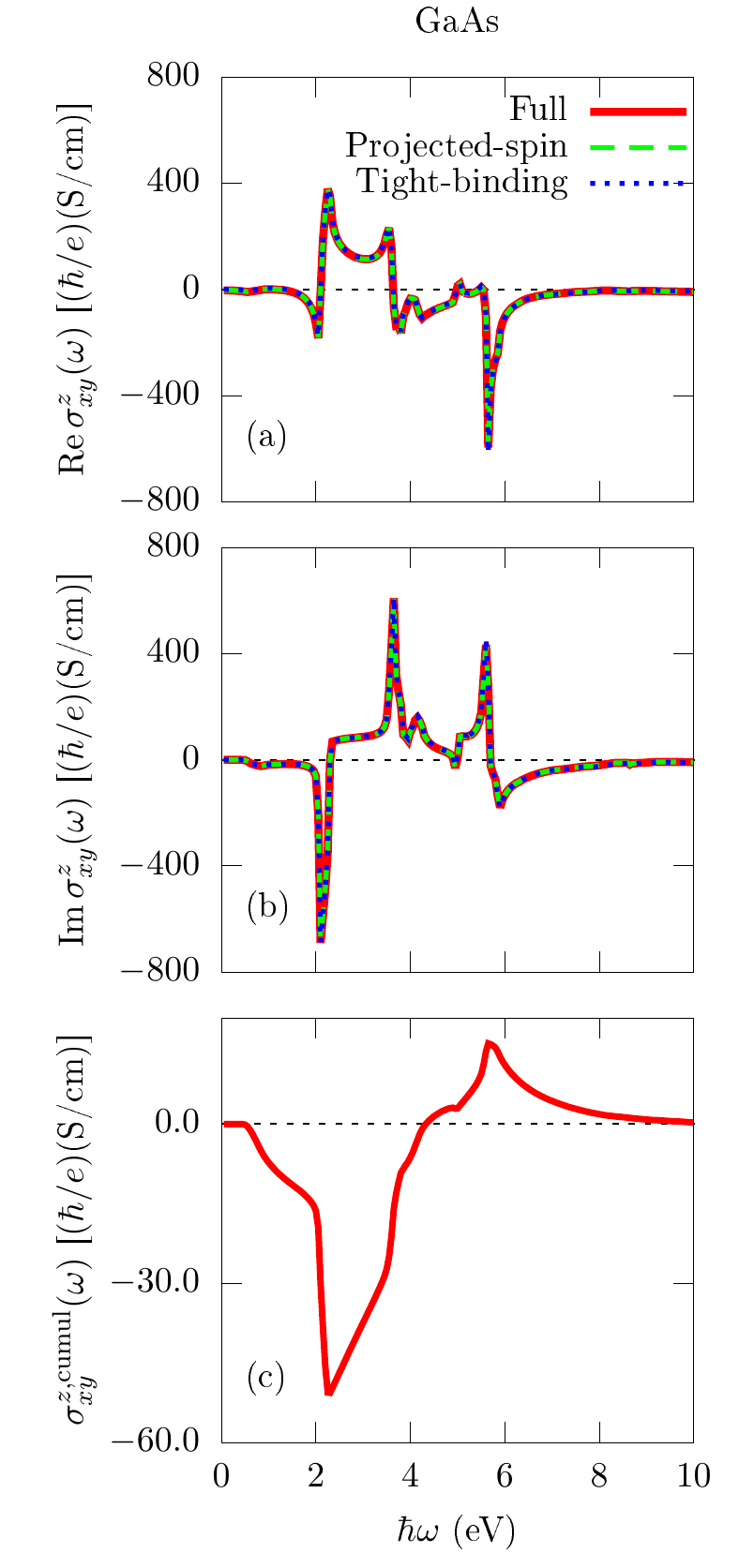}
\caption{ The (a) real and (b) imaginary parts of the dynamic SHC of
  bulk undoped GaAs.  (c) A cumulative integral [Eq.~\eqref{eq:cumul}]
  obtained from the imaginary part of the dynamic SHC.}
\label{fig:dyn_shc_GaAs}
\end{figure}
We conclude this section by discussing the calculated dynamic SHC of
undoped GaAs, shown in Figs.~\ref{fig:dyn_shc_GaAs}(a,b).
As in the case of the static SHC,
the results obtained with the
projected-spin and tight-binding approximations are very
close to the results of a full calculation.  To check the
convergence of the static SHC with respect to
$N_{\rm W}$ (or equivalently, to the
energy range spanned by the MLWFs), we calculate the
following function from the imaginary part of the dynamic SHC:
\begin{equation}
\sigma^{z,\mathrm{cumul}}_{xy}(\omega) = \frac{2}{\pi} \int_0^{\omega} {d\omega' \frac{\mathrm{Im}\, {\sigma_{xy}^{z}(\omega')}}{\omega'}}\,.
\label{eq:cumul}    
\end{equation}
Thanks to the Kramers-Kronig relation, this function converges
to the static SHC value (which is zero for undoped GaAs) as
$\omega \rightarrow \infty$.
As show in Fig.~\ref{fig:dyn_shc_GaAs}(c),
$\sigma_{xy}^{z,\mathrm{cumul}}(\omega)$ is already reasonably close
to zero by the time $\omega$ reaches 8~eV. This implies that,
for the purpose of evaluating the static
SHC, one should choose a set of MLWFs that
describes accurately the conduction bands of GaAs up to at least 8~eV
from the valence-band maximum.

In closing, we note that the dynamic SHC of GaAs
  reported in Ref.~\onlinecite{Guo2005PRL} is blue-shifted by around
  1~eV compared to our results in Figs.~\ref{fig:dyn_shc_GaAs}(a,b).
  Apart from that, the spectra obtained in both works
  are quite similar in terms of the overall shape, including the
  positions of the peaks. The reason for the blue shift is that the authors of
  Ref.~\onlinecite{Guo2005PRL} applied the so-called scissor shift in
  order to correct for the underestimation of the band gap in density-functional theory calculations.  We chose not to perform the
  scissors operation, in order to demonstrate in a transparent manner the Kramers-Kronig relation between the static SHC and the imaginary
  part of the dynamic SHC.

\section{Computational benefits of the approximations to the Wannier-interpolation scheme}
\label{sec:comp_cost}

\begin{figure}
  \includegraphics[width=0.95\columnwidth]{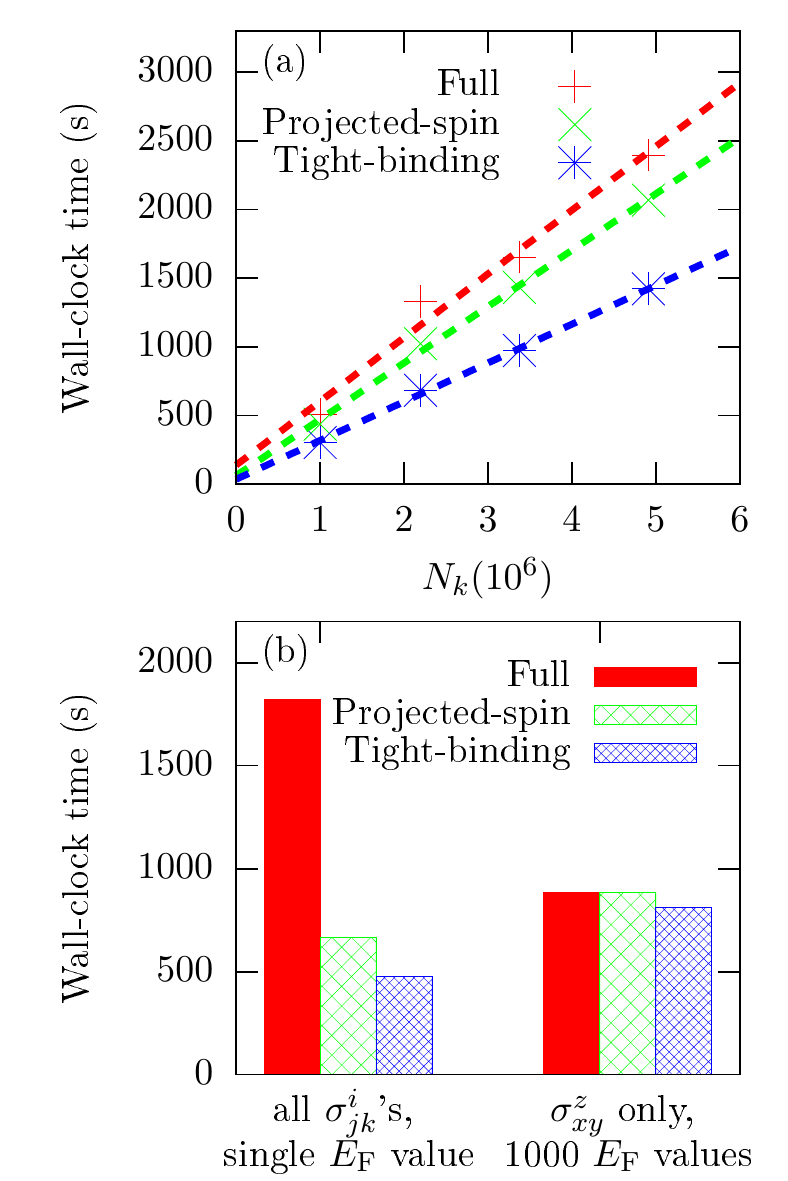}
  \caption{Timings of various benchmark calculations of the static SHC of fcc Pt,
  comparing the full Wannier-interpolation scheme proposed in this work with
  the two approximate schemes described in Sec.~\ref{sec:approx}.
  (a) The wall-clock time
 plotted as a function of the number $N_k$ of $k$ points on the interpolation grid.
 The dashed lines are a guide to the eye.
  (b) The wall-clock time for $N_k=10^6$, when all 27 components of SHC tensor are calculated independently for the unshifted Fermi level (left), and when only $\sigma^z_{xy}$ is calculated for 1000 different values of the Fermi level (right).
  }
  \label{fig:time}
\end{figure}
 
In this section, we analyze the gains in numerical efficiency from making
the projected-spin and tight-binding approximations when evaluating the SHC.
To that end we have measured, for the case of Pt with an unshifted Fermi level, 
the time elapsed in the subroutine that calculates the static SHC
$\sigma^z_{xy}(0)$. In all the calculations reported below the workload was distributed across 16 CPUs, with
parallelization over the $k$ points of the interpolation grid.

A full (approximation-free) Wannier-interpolation calculation of $\sigma^z_{xy}(0)$ requires seven
$N_{\rm W}\times N_{\rm W}$ matrices at each $k$ point,
namely,
\begin{subequations}
\begin{align}
\label{eq:matel-a}
&\matel{u_{m\mathbf{k}}^{{\rm (W)}}}{H_\mathbf{k}}{u_{n\mathbf{k}}^{{\rm (W)}}}\,,\\
\label{eq:matel-b}
\partial_j&\matel{u_{m\mathbf{k}}^{{\rm (W)}}}{H_\mathbf{k}}{u_{n\mathbf{k}}^{{\rm (W)}}}
\text{ with $j=x,y$}\,,\\
\label{eq:matel-c}
&\braket{u_{m\mathbf{k}}^{{\rm (W)}}}{\partial_y u_{n\mathbf{k}}^{{\rm (W)}}}\,,\\
\label{eq:matel-d}
&\matel{u_{m\mathbf{k}}^{\rm (W)}}{s^z}{u_{n\mathbf{k}}^{{\rm (W)}}}\,,\\
\label{eq:matel-e}
&\matel{u_{m\mathbf{k}}^{\rm (W)}}{s^z}{\partial_x u_{n\mathbf{k}}^{\rm (W)}}\,,\\
\label{eq:matel-f}
&\matel{u_{m\mathbf{k}}^{\rm (W)}}{s^z H_\mathbf{k}}{\partial_x
u_{n\mathbf{k}}^{\rm (W)}}\,.
\end{align}
\end{subequations}
These matrices are evaluated by performing Fourier transforms
from real space to $k$ space; the total time spent on
those Fourier transforms scales linearly with the number $N_k$ of $k$ points, 
and they constitute the most time-consuming step in the entire calculation.
On the other hand, a calculation using the projected-spin approximation requires six 
matrices: those in
Eqs.~\eqref{eq:matel-a}--\eqref{eq:matel-d}, as well as $\braket{u_{m\mathbf{k}}^{{\rm (W)}}}{\partial_x u_{n\mathbf{k}}^{{\rm (W)}}}$.
Finally, when making the tight-binding approximation
only the four matrices in Eqs.~\eqref{eq:matel-a}, \eqref{eq:matel-b}, and \eqref{eq:matel-d} are needed.

Figure~\ref{fig:time}(a) shows that in all three cases
the wall-clock time indeed scales linearly with $N_k$.
Compared to a tight-binding calculation,
a full calculation takes 1.7 times as long,
and a projected-spin calculation takes 1.4 times as long.
These results are consistent with the fact that
the projected-spin and full calculations require respectively
$6/4=1.5$ times and $7/4=1.75$ times more matrix elements
than the tight-binding calculation.
Note that the speed-up from the projected-spin approximation alone
is quite modest, around 15\%.

In low-symmetry crystals, the SHC tensor
can have
up to 27 independent components. To illustrate this scenario, we have
calculated explicitly all 27 components for Pt, disregarding the fact that
some are related by symmetry (for example, $\sigma^z_{xy}=\sigma^x_{yz}$).
The full calculation took
3.8 times as long as the tight-binding one, and 2.7 times  long as  the projected-spin one
[see Fig.~\ref{fig:time}(b)].
To understand these numbers, note that the full calculation requires 28 matrices of size $N_{\rm W}\times N_{\rm W}$ at each ${\bf k}$, namely,
\begin{eqnarray}
\matel{u_{m\mathbf{k}}^{(\textrm{W})}}{H_\mathbf{k}}{u_{n\mathbf{k}}^{(\textrm{W})}},\,
\partial_j\matel{u_{m\mathbf{k}}^{(\textrm{W})}}{H_\mathbf{k}}{u_{n\mathbf{k}}^{(\textrm{W})}},\,
\braket{u_{m\mathbf{k}}^{(\textrm{W})}}{\partial_j u_{m\mathbf{k}}^{(\textrm{W})}},\nonumber\\
\matel{u_{m\mathbf{k}}^{(\textrm{W})}}{s^i}{u_{n\mathbf{k}}^{(\textrm{W})}},\,
\matel{u_{m\mathbf{k}}^{(\textrm{W})}}{s^i}{\partial_j u_{n\mathbf{k}}^{(\textrm{W})}},\,
\matel{u_{m\mathbf{k}}^{(\textrm{W})}}{s^i H_\mathbf{k}}{\partial_j u_{n\mathbf{k}}^{(\textrm{W})}}.\nonumber\\
\end{eqnarray}
In comparison, the projected-spin and tight-binding calculations requires ten and seven matrices, respectively.
This yields estimated speed-ups of 28/10 = 2.8 and 28/7 = 4, respectively, which are quite consistent with the actual speed-ups quoted above.

Consider now a scenario where the static SHC $\sigma^z_{xy}(0)$ is evaluated
for a large number of Fermi levels
(as in Sec.~\ref{sec:shc_results}), or where the dynamic SHC $\sigma^z_{xy}(\omega)$ is evaluated
for a large number of frequencies
(as in Sec.~\ref{sec:dyn_shc_results}).
In both cases, the time spent interpolating the $k$-space matrix elements becomes
small compared to the time spent evaluating the Kubo formula 
starting from those matrix elements.
For example, calculating $\sigma^z_{xy}(0)$ 
on a $100 \times 100 \times 100$ $k$-point mesh for 1000 Fermi levels
 takes 880 seconds
with the full interpolation approach,
and 810 seconds with the tight-binding
approximation [see Fig.~\ref{fig:time} (b)].
This marginal increase by 8\% in efficiency is offset by a loss
of accuracy of about 10\%. Therefore, in these usage scenarios it is best
to perform the full calculation, without further approximations to the matrix elements.

From this series of benchmark calculations, we conclude that the speed-up obtained by making
the tight-binding approximation in
the Wannier-interpolation scheme
varies from marginal to up to a factor of four (depending on
the number of independent components of the SHC tensor, and on
the number of Fermi-energy or frequency values).
The main virtue of
the tight-binding approach, however, is that the needed parameters can
be obtained by fitting to a first-principles calculation or to experiment,
and thus it significantly lowers the barrier for computing the SHC with reasonable accuracy (see Sec.~\ref{sec:shc_results}).

\section{Summary}
\label{sec:conclusion}

In summary, we have presented a  method
to calculate the intrinsic SHC of a crystal by interpolating
the velocity and spin-current matrix elements using MLWFs.
The interpolation is carried out as a post-processing step following
a conventional first-principles calculation, taking as
input from the {\it ab initio} run the matrix elements 
\begin{align}
&\left<u_{m{\bf q}} \right|H_{\bf q} \left|
u_{n{\bf q}} \right>=\varepsilon_{m{\bf q}}\delta_{mn} \,,
\left<u_{m{\bf q}} \middle|
u_{n,{\bf q}+{\bf b}} \right> \,,
\left<u_{m{\bf q}} \right|s^i \left|
u_{n{\bf q}} \right> \,,
\nonumber\\
&\left<u_{m{\bf q}} \right|s^i
\left|u_{n,{\bf q}+{\bf b}} \right> \,,
\left<u_{m{\bf q}} \right| s^i H_{\bf q}
\left| u_{n, {\bf q}+{\bf b}}\right> 
\label{eq:q-list}
\end{align}
between the set of Bloch eigenstates used for generating the MLWFs.
The method was validated
by applying it to fcc Pt and zincblende GaAs and comparing
the results with previous {\it ab initio} calculations for these materials. 

In addition, we have
considered two different types of approximations that have been
frequently used in previous Wannier-based calculations of the SHC:
the projected-spin approximation,
and the tight-binding approximation.
We found that the projected-spin approximation
is quite good for both Pt and GaAs, and that at worst it slows down
the convergence of the calculated SHC
with respect to the number of Wannier functions.
As for the tight-binding approximation (which includes the
projected-spin approximation),
it introduces errors of the order of 10\%.
This suggests that
empirical tight-binding models,
with parameters fitted to  experimental results or to first-principles calculations,
can yield reasonably accurate values for the SHC.

{\it Note added:} While we were finalizing this study, we became aware
of a related preprint, now published as
Ref.~\onlinecite{qiao2018PRB}.  Although the notation in
Ref.~\onlinecite{qiao2018PRB} is different from the one used in the
present work, our proposed method is broadly consistent with the one reported
therein. The viewpoints and analyses of the two studies are however substantially different.
One noteworthy difference is
the way in which the last two matrix elements in Eq.~\eqref{eq:q-list} are evaluated. Here we calculate them exactly,
whereas in Ref.~\onlinecite{qiao2018PRB} they are expressed
 in terms of the first three by making the approximations
\begin{eqnarray}
&&\left< u_{m{\bf q}}\right| s^i
\left| u_{n,{\bf q}+{\bf b}}\right>\nonumber\\
&&\approx\sum_{l}\,
\left< u_{m{\bf q}}\right| s^i \left| u_{l{\bf q}} \right>
\left< u_{l{\bf q}}\right|
\left. u_{n,{\bf q}+{\bf b}}\right>
\label{eq:qiao_1}
\end{eqnarray}
and
\begin{eqnarray}
&&\left< u_{m{\bf q}}\right| s^i H_\mathbf{q}
\left| u_{n,{\bf q}+{\bf b}}\right>\nonumber\\
&&\approx\sum_{l}\,
\left< u_{m{\bf q}}\right| s^i \left| u_{l{\bf q}} \right>
\varepsilon_{l{\bf q}}
\left< u_{l{\bf q}}\right|
\left. u_{n,{\bf q}+{\bf b}}\right>\,,
\label{eq:qiao_H}
\end{eqnarray}
where the index $l$ runs
over the {\it ab initio} Bloch eigenstates at the grid point ${\bf q}$ that were included  when
constructing the MLWFs. This 
is similar in spirit to (but more accurate than) the projected-spin approximation
of Eq.~\eqref{eq:svW_approx}.

\begin{acknowledgments}
J.H.R. thanks the organizers of
E-CAM Wannier90 Software Development Workshop held in September 2016
during which part of this work was done
and the hospitality of Centro de F\'isica de Materiales,
Universidad del Pa\'is Vasco, San Sebasti\'an.
This work was supported by the Creative-Pioneering Research Program through Seoul
National University and Korean NRF No-2016R1A1A1A05919979 (J.H.R. and C.-H.P.),
and by Grant No. FIS2016-77188-P from the Spanish Ministerio de Econom\'ia
y Competitividad (I.S.).
Computational resources were provided by KISTI Supercomputing Center
(KSC-2018-CHA-0051).
\end{acknowledgments}

\bibliography{main}

\begin{thebibliography}{34}%
\makeatletter
\providecommand \@ifxundefined [1]{%
 \@ifx{#1\undefined}
}%
\providecommand \@ifnum [1]{%
 \ifnum #1\expandafter \@firstoftwo
 \else \expandafter \@secondoftwo
 \fi
}%
\providecommand \@ifx [1]{%
 \ifx #1\expandafter \@firstoftwo
 \else \expandafter \@secondoftwo
 \fi
}%
\providecommand \natexlab [1]{#1}%
\providecommand \enquote  [1]{``#1''}%
\providecommand \bibnamefont  [1]{#1}%
\providecommand \bibfnamefont [1]{#1}%
\providecommand \citenamefont [1]{#1}%
\providecommand \href@noop [0]{\@secondoftwo}%
\providecommand \href [0]{\begingroup \@sanitize@url \@href}%
\providecommand \@href[1]{\@@startlink{#1}\@@href}%
\providecommand \@@href[1]{\endgroup#1\@@endlink}%
\providecommand \@sanitize@url [0]{\catcode `\\12\catcode `\$12\catcode
  `\&12\catcode `\#12\catcode `\^12\catcode `\_12\catcode `\%12\relax}%
\providecommand \@@startlink[1]{}%
\providecommand \@@endlink[0]{}%
\providecommand \url  [0]{\begingroup\@sanitize@url \@url }%
\providecommand \@url [1]{\endgroup\@href {#1}{\urlprefix }}%
\providecommand \urlprefix  [0]{URL }%
\providecommand \Eprint [0]{\href }%
\providecommand \doibase [0]{http://dx.doi.org/}%
\providecommand \selectlanguage [0]{\@gobble}%
\providecommand \bibinfo  [0]{\@secondoftwo}%
\providecommand \bibfield  [0]{\@secondoftwo}%
\providecommand \translation [1]{[#1]}%
\providecommand \BibitemOpen [0]{}%
\providecommand \bibitemStop [0]{}%
\providecommand \bibitemNoStop [0]{.\EOS\space}%
\providecommand \EOS [0]{\spacefactor3000\relax}%
\providecommand \BibitemShut  [1]{\csname bibitem#1\endcsname}%
\let\auto@bib@innerbib\@empty
\bibitem [{\citenamefont {Sinova}\ \emph {et~al.}(2015)\citenamefont {Sinova},
  \citenamefont {Valenzuela}, \citenamefont {Wunderlich}, \citenamefont
  {Back},\ and\ \citenamefont {Jungwirth}}]{sinova2015RMP}%
  \BibitemOpen
  \bibfield  {author} {\bibinfo {author} {\bibfnamefont {J.}~\bibnamefont
  {Sinova}}, \bibinfo {author} {\bibfnamefont {S.~O.}\ \bibnamefont
  {Valenzuela}}, \bibinfo {author} {\bibfnamefont {J.}~\bibnamefont
  {Wunderlich}}, \bibinfo {author} {\bibfnamefont {C.H.}\ \bibnamefont {Back}},
  \ and\ \bibinfo {author} {\bibfnamefont {T.}~\bibnamefont {Jungwirth}},\
  }\bibfield  {title} {\enquote {\bibinfo {title} {{Spin Hall effects}},}\
  }\href {\doibase 10.1103/RevModPhys.87.1213} {\bibfield  {journal} {\bibinfo
  {journal} {Rev. Mod. Phys.}\ }\textbf {\bibinfo {volume} {87}},\ \bibinfo
  {pages} {1213} (\bibinfo {year} {2015})}\BibitemShut {NoStop}%
\bibitem [{\citenamefont {Sinova}\ and\ \citenamefont
  {{\v{Z}}uti{\'c}}(2012)}]{sinova2012NatMater}%
  \BibitemOpen
  \bibfield  {author} {\bibinfo {author} {\bibfnamefont {J.}~\bibnamefont
  {Sinova}}\ and\ \bibinfo {author} {\bibfnamefont {I.}~\bibnamefont
  {{\v{Z}}uti{\'c}}},\ }\bibfield  {title} {\enquote {\bibinfo {title} {{New
  moves of the spintronics tango}},}\ }\href {\doibase 10.1038/nmat3304}
  {\bibfield  {journal} {\bibinfo  {journal} {Nat. Mater.}\ }\textbf {\bibinfo
  {volume} {11}},\ \bibinfo {pages} {368} (\bibinfo {year} {2012})}\BibitemShut
  {NoStop}%
\bibitem [{\citenamefont {Jungwirth}\ \emph {et~al.}(2012)\citenamefont
  {Jungwirth}, \citenamefont {Wunderlich},\ and\ \citenamefont
  {Olejn{\'\i}k}}]{jungwirth2012NatMater}%
  \BibitemOpen
  \bibfield  {author} {\bibinfo {author} {\bibfnamefont {T.}~\bibnamefont
  {Jungwirth}}, \bibinfo {author} {\bibfnamefont {J.}~\bibnamefont
  {Wunderlich}}, \ and\ \bibinfo {author} {\bibfnamefont {K.}~\bibnamefont
  {Olejn{\'\i}k}},\ }\bibfield  {title} {\enquote {\bibinfo {title} {{Spin Hall
  effect devices}},}\ }\href {\doibase doi.org/10.1038/nmat3279} {\bibfield
  {journal} {\bibinfo  {journal} {Nat. Mater.}\ }\textbf {\bibinfo {volume}
  {11}},\ \bibinfo {pages} {382} (\bibinfo {year} {2012})}\BibitemShut
  {NoStop}%
\bibitem [{\citenamefont {Hirsch}(1999)}]{Hirsch1999PRL}%
  \BibitemOpen
  \bibfield  {author} {\bibinfo {author} {\bibfnamefont {J.~E.}\ \bibnamefont
  {Hirsch}},\ }\bibfield  {title} {\enquote {\bibinfo {title} {{Spin Hall
  Effect}},}\ }\href {\doibase 10.1103/PhysRevLett.83.1834} {\bibfield
  {journal} {\bibinfo  {journal} {Phys. Rev. Lett.}\ }\textbf {\bibinfo
  {volume} {83}},\ \bibinfo {pages} {1834} (\bibinfo {year}
  {1999})}\BibitemShut {NoStop}%
\bibitem [{\citenamefont {Saitoh}\ \emph {et~al.}(2006)\citenamefont {Saitoh},
  \citenamefont {Ueda}, \citenamefont {Miyajima},\ and\ \citenamefont
  {Tatara}}]{Saitoh2006APL}%
  \BibitemOpen
  \bibfield  {author} {\bibinfo {author} {\bibfnamefont {E.}~\bibnamefont
  {Saitoh}}, \bibinfo {author} {\bibfnamefont {M.}~\bibnamefont {Ueda}},
  \bibinfo {author} {\bibfnamefont {H.}~\bibnamefont {Miyajima}}, \ and\
  \bibinfo {author} {\bibfnamefont {G.}~\bibnamefont {Tatara}},\ }\bibfield
  {title} {\enquote {\bibinfo {title} {{Conversion of spin current into charge
  current at room temperature: Inverse spin-Hall effect}},}\ }\href {\doibase
  10.1063/1.2199473} {\bibfield  {journal} {\bibinfo  {journal} {Appl. Phys.
  Lett.}\ }\textbf {\bibinfo {volume} {88}},\ \bibinfo {pages} {182509}
  (\bibinfo {year} {2006})}\BibitemShut {NoStop}%
\bibitem [{\citenamefont {Kimura}\ \emph {et~al.}(2007)\citenamefont {Kimura},
  \citenamefont {Otani}, \citenamefont {Sato}, \citenamefont {Takahashi},\ and\
  \citenamefont {Maekawa}}]{Kimura2007PRL}%
  \BibitemOpen
  \bibfield  {author} {\bibinfo {author} {\bibfnamefont {T.}~\bibnamefont
  {Kimura}}, \bibinfo {author} {\bibfnamefont {Y.}~\bibnamefont {Otani}},
  \bibinfo {author} {\bibfnamefont {T.}~\bibnamefont {Sato}}, \bibinfo {author}
  {\bibfnamefont {S.}~\bibnamefont {Takahashi}}, \ and\ \bibinfo {author}
  {\bibfnamefont {S.}~\bibnamefont {Maekawa}},\ }\bibfield  {title} {\enquote
  {\bibinfo {title} {{Room-Temperature Reversible Spin Hall Effect}},}\ }\href
  {\doibase 10.1103/PhysRevLett.98.156601} {\bibfield  {journal} {\bibinfo
  {journal} {Phys. Rev. Lett.}\ }\textbf {\bibinfo {volume} {98}},\ \bibinfo
  {pages} {156601} (\bibinfo {year} {2007})}\BibitemShut {NoStop}%
\bibitem [{\citenamefont {Wunderlich}\ \emph {et~al.}(2010)\citenamefont
  {Wunderlich}, \citenamefont {Park}, \citenamefont {Irvine}, \citenamefont
  {Z{\^a}rbo}, \citenamefont {Rozkotov{\'a}}, \citenamefont {Nemec},
  \citenamefont {Nov{\'a}k}, \citenamefont {Sinova},\ and\ \citenamefont
  {Jungwirth}}]{WunderlichScience2010}%
  \BibitemOpen
  \bibfield  {author} {\bibinfo {author} {\bibfnamefont {J.}~\bibnamefont
  {Wunderlich}}, \bibinfo {author} {\bibfnamefont {B.-G.}\ \bibnamefont
  {Park}}, \bibinfo {author} {\bibfnamefont {A.~C.}\ \bibnamefont {Irvine}},
  \bibinfo {author} {\bibfnamefont {L.~P.}\ \bibnamefont {Z{\^a}rbo}}, \bibinfo
  {author} {\bibfnamefont {E.}~\bibnamefont {Rozkotov{\'a}}}, \bibinfo {author}
  {\bibfnamefont {P.}~\bibnamefont {Nemec}}, \bibinfo {author} {\bibfnamefont
  {V.}~\bibnamefont {Nov{\'a}k}}, \bibinfo {author} {\bibfnamefont
  {J.}~\bibnamefont {Sinova}}, \ and\ \bibinfo {author} {\bibfnamefont
  {T.}~\bibnamefont {Jungwirth}},\ }\bibfield  {title} {\enquote {\bibinfo
  {title} {{Spin Hall Effect Transistor}},}\ }\href {\doibase
  10.1126/science.1195816} {\bibfield  {journal} {\bibinfo  {journal}
  {Science}\ }\textbf {\bibinfo {volume} {330}},\ \bibinfo {pages} {1801}
  (\bibinfo {year} {2010})}\BibitemShut {NoStop}%
\bibitem [{\citenamefont {Ando}\ \emph {et~al.}(2008)\citenamefont {Ando},
  \citenamefont {Takahashi}, \citenamefont {Harii}, \citenamefont {Sasage},
  \citenamefont {Ieda}, \citenamefont {Maekawa},\ and\ \citenamefont
  {Saitoh}}]{Ando2008PRL}%
  \BibitemOpen
  \bibfield  {author} {\bibinfo {author} {\bibfnamefont {K.}~\bibnamefont
  {Ando}}, \bibinfo {author} {\bibfnamefont {S.}~\bibnamefont {Takahashi}},
  \bibinfo {author} {\bibfnamefont {K.}~\bibnamefont {Harii}}, \bibinfo
  {author} {\bibfnamefont {K.}~\bibnamefont {Sasage}}, \bibinfo {author}
  {\bibfnamefont {J.}~\bibnamefont {Ieda}}, \bibinfo {author} {\bibfnamefont
  {S.}~\bibnamefont {Maekawa}}, \ and\ \bibinfo {author} {\bibfnamefont
  {E.}~\bibnamefont {Saitoh}},\ }\bibfield  {title} {\enquote {\bibinfo {title}
  {{Electric Manipulation of Spin Relaxation Using the Spin Hall Effect}},}\
  }\href {\doibase 10.1103/PhysRevLett.101.036601} {\bibfield  {journal}
  {\bibinfo  {journal} {Phys. Rev. Lett.}\ }\textbf {\bibinfo {volume} {101}},\
  \bibinfo {pages} {036601} (\bibinfo {year} {2008})}\BibitemShut {NoStop}%
\bibitem [{\citenamefont {Liu}\ \emph {et~al.}(2011)\citenamefont {Liu},
  \citenamefont {Moriyama}, \citenamefont {Ralph},\ and\ \citenamefont
  {Buhrman}}]{Liu2011PRL}%
  \BibitemOpen
  \bibfield  {author} {\bibinfo {author} {\bibfnamefont {Luqiao}\ \bibnamefont
  {Liu}}, \bibinfo {author} {\bibfnamefont {Takahiro}\ \bibnamefont
  {Moriyama}}, \bibinfo {author} {\bibfnamefont {D.~C.}\ \bibnamefont {Ralph}},
  \ and\ \bibinfo {author} {\bibfnamefont {R.~A.}\ \bibnamefont {Buhrman}},\
  }\bibfield  {title} {\enquote {\bibinfo {title} {{Spin-Torque Ferromagnetic
  Resonance Induced by the Spin Hall Effect}},}\ }\href {\doibase
  10.1103/PhysRevLett.106.036601} {\bibfield  {journal} {\bibinfo  {journal}
  {Phys. Rev. Lett.}\ }\textbf {\bibinfo {volume} {106}},\ \bibinfo {pages}
  {036601} (\bibinfo {year} {2011})}\BibitemShut {NoStop}%
\bibitem [{\citenamefont {Liu}\ \emph {et~al.}(2012)\citenamefont {Liu},
  \citenamefont {Pai}, \citenamefont {Li}, \citenamefont {Tseng}, \citenamefont
  {Ralph},\ and\ \citenamefont {Buhrman}}]{Liu2012Science}%
  \BibitemOpen
  \bibfield  {author} {\bibinfo {author} {\bibfnamefont {L.}~\bibnamefont
  {Liu}}, \bibinfo {author} {\bibfnamefont {C.-F.}\ \bibnamefont {Pai}},
  \bibinfo {author} {\bibfnamefont {Y.}~\bibnamefont {Li}}, \bibinfo {author}
  {\bibfnamefont {H.~W.}\ \bibnamefont {Tseng}}, \bibinfo {author}
  {\bibfnamefont {D.~C.}\ \bibnamefont {Ralph}}, \ and\ \bibinfo {author}
  {\bibfnamefont {R.~A.}\ \bibnamefont {Buhrman}},\ }\bibfield  {title}
  {\enquote {\bibinfo {title} {{Spin-Torque Switching with the Giant Spin Hall
  Effect of Tantalum}},}\ }\href {\doibase 10.1126/science.1218197} {\bibfield
  {journal} {\bibinfo  {journal} {Science}\ }\textbf {\bibinfo {volume}
  {336}},\ \bibinfo {pages} {555} (\bibinfo {year} {2012})}\BibitemShut
  {NoStop}%
\bibitem [{\citenamefont {Nagaosa}\ \emph {et~al.}(2010)\citenamefont
  {Nagaosa}, \citenamefont {Sinova}, \citenamefont {Onoda}, \citenamefont
  {MacDonald},\ and\ \citenamefont {Ong}}]{nagaosa2010RMP}%
  \BibitemOpen
  \bibfield  {author} {\bibinfo {author} {\bibfnamefont {N.}~\bibnamefont
  {Nagaosa}}, \bibinfo {author} {\bibfnamefont {J.}~\bibnamefont {Sinova}},
  \bibinfo {author} {\bibfnamefont {S.}~\bibnamefont {Onoda}}, \bibinfo
  {author} {\bibfnamefont {A.H.}\ \bibnamefont {MacDonald}}, \ and\ \bibinfo
  {author} {\bibfnamefont {N.P.}\ \bibnamefont {Ong}},\ }\bibfield  {title}
  {\enquote {\bibinfo {title} {{Anomalous Hall effect}},}\ }\href {\doibase
  10.1103/RevModPhys.82.1539} {\bibfield  {journal} {\bibinfo  {journal} {Rev.
  Mod. Phys.}\ }\textbf {\bibinfo {volume} {82}},\ \bibinfo {pages} {1539}
  (\bibinfo {year} {2010})}\BibitemShut {NoStop}%
\bibitem [{\citenamefont {Inoue}\ \emph {et~al.}(2004)\citenamefont {Inoue},
  \citenamefont {Bauer},\ and\ \citenamefont {Molenkamp}}]{inoue2004PRB}%
  \BibitemOpen
  \bibfield  {author} {\bibinfo {author} {\bibfnamefont {J.}~\bibnamefont
  {Inoue}}, \bibinfo {author} {\bibfnamefont {G.~E.~W.}\ \bibnamefont {Bauer}},
  \ and\ \bibinfo {author} {\bibfnamefont {L.~W.}\ \bibnamefont {Molenkamp}},\
  }\bibfield  {title} {\enquote {\bibinfo {title} {{Suppression of the
  persistent spin Hall current by defect scattering}},}\ }\href {\doibase
  10.1103/PhysRevB.70.041303} {\bibfield  {journal} {\bibinfo  {journal} {Phys.
  Rev. B}\ }\textbf {\bibinfo {volume} {70}},\ \bibinfo {pages} {041303}
  (\bibinfo {year} {2004})}\BibitemShut {NoStop}%
\bibitem [{\citenamefont {Bernevig}\ and\ \citenamefont
  {Zhang}(2005)}]{bernevig2005PRL}%
  \BibitemOpen
  \bibfield  {author} {\bibinfo {author} {\bibfnamefont {B.~A.}\ \bibnamefont
  {Bernevig}}\ and\ \bibinfo {author} {\bibfnamefont {S.-C.}\ \bibnamefont
  {Zhang}},\ }\bibfield  {title} {\enquote {\bibinfo {title} {{Intrinsic Spin
  Hall Effect in the Two-Dimensional Hole Gas}},}\ }\href {\doibase
  10.1103/PhysRevLett.95.016801} {\bibfield  {journal} {\bibinfo  {journal}
  {Phys. Rev. Lett.}\ }\textbf {\bibinfo {volume} {95}},\ \bibinfo {pages}
  {016801} (\bibinfo {year} {2005})}\BibitemShut {NoStop}%
\bibitem [{\citenamefont {Chadova}\ \emph {et~al.}(2015)\citenamefont
  {Chadova}, \citenamefont {Fedorov}, \citenamefont {Herschbach}, \citenamefont
  {Gradhand}, \citenamefont {Mertig}, \citenamefont {K{\"o}dderitzsch},\ and\
  \citenamefont {Ebert}}]{chadova2015PRB}%
  \BibitemOpen
  \bibfield  {author} {\bibinfo {author} {\bibfnamefont {K.}~\bibnamefont
  {Chadova}}, \bibinfo {author} {\bibfnamefont {D.~V.}\ \bibnamefont
  {Fedorov}}, \bibinfo {author} {\bibfnamefont {C.}~\bibnamefont {Herschbach}},
  \bibinfo {author} {\bibfnamefont {M.}~\bibnamefont {Gradhand}}, \bibinfo
  {author} {\bibfnamefont {I.}~\bibnamefont {Mertig}}, \bibinfo {author}
  {\bibfnamefont {D.}~\bibnamefont {K{\"o}dderitzsch}}, \ and\ \bibinfo
  {author} {\bibfnamefont {H.}~\bibnamefont {Ebert}},\ }\bibfield  {title}
  {\enquote {\bibinfo {title} {{Separation of the individual contributions to
  the spin Hall effect in dilute alloys within the first-principles
  Kubo-St{\v{r}}eda approach}},}\ }\href {\doibase 10.1103/PhysRevB.92.045120}
  {\bibfield  {journal} {\bibinfo  {journal} {Phys. Rev. B}\ }\textbf {\bibinfo
  {volume} {92}},\ \bibinfo {pages} {045120} (\bibinfo {year}
  {2015})}\BibitemShut {NoStop}%
\bibitem [{\citenamefont {Guo}\ \emph {et~al.}(2008)\citenamefont {Guo},
  \citenamefont {Murakami}, \citenamefont {Chen},\ and\ \citenamefont
  {Nagaosa}}]{Guo2008PRL}%
  \BibitemOpen
  \bibfield  {author} {\bibinfo {author} {\bibfnamefont {G.~Y.}\ \bibnamefont
  {Guo}}, \bibinfo {author} {\bibfnamefont {S.}~\bibnamefont {Murakami}},
  \bibinfo {author} {\bibfnamefont {T.-W.}\ \bibnamefont {Chen}}, \ and\
  \bibinfo {author} {\bibfnamefont {N.}~\bibnamefont {Nagaosa}},\ }\bibfield
  {title} {\enquote {\bibinfo {title} {{Intrinsic Spin Hall Effect in Platinum:
  First-Principles Calculations}},}\ }\href {\doibase
  10.1103/PhysRevLett.100.096401} {\bibfield  {journal} {\bibinfo  {journal}
  {Phys. Rev. Lett.}\ }\textbf {\bibinfo {volume} {100}},\ \bibinfo {pages}
  {096401} (\bibinfo {year} {2008})}\BibitemShut {NoStop}%
\bibitem [{\citenamefont {Guo}\ \emph {et~al.}(2005)\citenamefont {Guo},
  \citenamefont {Yao},\ and\ \citenamefont {Niu}}]{Guo2005PRL}%
  \BibitemOpen
  \bibfield  {author} {\bibinfo {author} {\bibfnamefont {G.~Y.}\ \bibnamefont
  {Guo}}, \bibinfo {author} {\bibfnamefont {Y.}~\bibnamefont {Yao}}, \ and\
  \bibinfo {author} {\bibfnamefont {Q.}~\bibnamefont {Niu}},\ }\bibfield
  {title} {\enquote {\bibinfo {title} {{Ab initio Calculation of the Intrinsic
  Spin Hall Effect in Semiconductors}},}\ }\href {\doibase
  10.1103/PhysRevLett.94.226601} {\bibfield  {journal} {\bibinfo  {journal}
  {Phys. Rev. Lett.}\ }\textbf {\bibinfo {volume} {94}},\ \bibinfo {pages}
  {226601} (\bibinfo {year} {2005})}\BibitemShut {NoStop}%
\bibitem [{\citenamefont {Guo}(2009)}]{Guo2009JAP}%
  \BibitemOpen
  \bibfield  {author} {\bibinfo {author} {\bibfnamefont {G.Y.}\ \bibnamefont
  {Guo}},\ }\bibfield  {title} {\enquote {\bibinfo {title} {{Ab initio
  calculation of intrinsic spin Hall conductivity of Pd and Au}},}\ }\href
  {\doibase 10.1063/1.3054362} {\bibfield  {journal} {\bibinfo  {journal} {J.
  App. Phys.}\ }\textbf {\bibinfo {volume} {105}},\ \bibinfo {pages} {07C701}
  (\bibinfo {year} {2009})}\BibitemShut {NoStop}%
\bibitem [{\citenamefont {Sui}\ \emph {et~al.}(2017)\citenamefont {Sui},
  \citenamefont {Wang}, \citenamefont {Kim}, \citenamefont {Wang},
  \citenamefont {Rhim}, \citenamefont {Duan},\ and\ \citenamefont
  {Kioussis}}]{Sui2017PRB}%
  \BibitemOpen
  \bibfield  {author} {\bibinfo {author} {\bibfnamefont {X.}~\bibnamefont
  {Sui}}, \bibinfo {author} {\bibfnamefont {C.}~\bibnamefont {Wang}}, \bibinfo
  {author} {\bibfnamefont {J.}~\bibnamefont {Kim}}, \bibinfo {author}
  {\bibfnamefont {J.}~\bibnamefont {Wang}}, \bibinfo {author} {\bibfnamefont
  {S.~H.}\ \bibnamefont {Rhim}}, \bibinfo {author} {\bibfnamefont
  {W.}~\bibnamefont {Duan}}, \ and\ \bibinfo {author} {\bibfnamefont
  {N.}~\bibnamefont {Kioussis}},\ }\bibfield  {title} {\enquote {\bibinfo
  {title} {{Giant enhancement of the intrinsic spin Hall conductivity in
  $\ensuremath{\beta}$-tungsten via substitutional doping}},}\ }\href {\doibase
  10.1103/PhysRevB.96.241105} {\bibfield  {journal} {\bibinfo  {journal} {Phys.
  Rev. B}\ }\textbf {\bibinfo {volume} {96}},\ \bibinfo {pages} {241105}
  (\bibinfo {year} {2017})}\BibitemShut {NoStop}%
\bibitem [{\citenamefont {Marzari}\ and\ \citenamefont
  {Vanderbilt}(1997)}]{Marzari1997PRB}%
  \BibitemOpen
  \bibfield  {author} {\bibinfo {author} {\bibfnamefont {N.}~\bibnamefont
  {Marzari}}\ and\ \bibinfo {author} {\bibfnamefont {D.}~\bibnamefont
  {Vanderbilt}},\ }\bibfield  {title} {\enquote {\bibinfo {title} {{Maximally
  localized generalized Wannier functions for composite energy bands}},}\
  }\href {\doibase 10.1103/PhysRevB.56.12847} {\bibfield  {journal} {\bibinfo
  {journal} {Phys. Rev. B}\ }\textbf {\bibinfo {volume} {56}},\ \bibinfo
  {pages} {12847} (\bibinfo {year} {1997})}\BibitemShut {NoStop}%
\bibitem [{\citenamefont {Souza}\ \emph {et~al.}(2001)\citenamefont {Souza},
  \citenamefont {Marzari},\ and\ \citenamefont {Vanderbilt}}]{Souza2001PRB}%
  \BibitemOpen
  \bibfield  {author} {\bibinfo {author} {\bibfnamefont {I.}~\bibnamefont
  {Souza}}, \bibinfo {author} {\bibfnamefont {N.}~\bibnamefont {Marzari}}, \
  and\ \bibinfo {author} {\bibfnamefont {D.}~\bibnamefont {Vanderbilt}},\
  }\bibfield  {title} {\enquote {\bibinfo {title} {{Maximally localized Wannier
  functions for entangled energy bands}},}\ }\href {\doibase
  10.1103/PhysRevB.65.035109} {\bibfield  {journal} {\bibinfo  {journal} {Phys.
  Rev. B}\ }\textbf {\bibinfo {volume} {65}},\ \bibinfo {pages} {035109}
  (\bibinfo {year} {2001})}\BibitemShut {NoStop}%
\bibitem [{\citenamefont {Mostofi}\ \emph {et~al.}(2008)\citenamefont
  {Mostofi}, \citenamefont {Yates}, \citenamefont {Lee}, \citenamefont {Souza},
  \citenamefont {Vanderbilt},\ and\ \citenamefont {Marzari}}]{mostofi2008}%
  \BibitemOpen
  \bibfield  {author} {\bibinfo {author} {\bibfnamefont {A.~A.}\ \bibnamefont
  {Mostofi}}, \bibinfo {author} {\bibfnamefont {J.~R.}\ \bibnamefont {Yates}},
  \bibinfo {author} {\bibfnamefont {Y.-S.}\ \bibnamefont {Lee}}, \bibinfo
  {author} {\bibfnamefont {I.}~\bibnamefont {Souza}}, \bibinfo {author}
  {\bibfnamefont {D.}~\bibnamefont {Vanderbilt}}, \ and\ \bibinfo {author}
  {\bibfnamefont {N.}~\bibnamefont {Marzari}},\ }\bibfield  {title} {\enquote
  {\bibinfo {title} {{wannier90: A tool for obtaining maximally-localised
  Wannier functions}},}\ }\href {\doibase 10.1016/j.cpc.2007.11.016} {\bibfield
   {journal} {\bibinfo  {journal} {Comput. Phys. Commun.}\ }\textbf {\bibinfo
  {volume} {178}},\ \bibinfo {pages} {685} (\bibinfo {year}
  {2008})}\BibitemShut {NoStop}%
\bibitem [{\citenamefont {Marzari}\ \emph {et~al.}(2012)\citenamefont
  {Marzari}, \citenamefont {Mostofi}, \citenamefont {Yates}, \citenamefont
  {Souza},\ and\ \citenamefont {Vanderbilt}}]{Marzari2012RMP}%
  \BibitemOpen
  \bibfield  {author} {\bibinfo {author} {\bibfnamefont {N.}~\bibnamefont
  {Marzari}}, \bibinfo {author} {\bibfnamefont {A.~A.}\ \bibnamefont
  {Mostofi}}, \bibinfo {author} {\bibfnamefont {J.~R.}\ \bibnamefont {Yates}},
  \bibinfo {author} {\bibfnamefont {I.}~\bibnamefont {Souza}}, \ and\ \bibinfo
  {author} {\bibfnamefont {D.}~\bibnamefont {Vanderbilt}},\ }\bibfield  {title}
  {\enquote {\bibinfo {title} {{Maximally localized Wannier functions: Theory
  and applications}},}\ }\href {\doibase 10.1103/RevModPhys.84.1419} {\bibfield
   {journal} {\bibinfo  {journal} {Rev. Mod. Phys.}\ }\textbf {\bibinfo
  {volume} {84}},\ \bibinfo {pages} {1419} (\bibinfo {year}
  {2012})}\BibitemShut {NoStop}%
\bibitem [{\citenamefont {Wang}\ \emph {et~al.}(2006)\citenamefont {Wang},
  \citenamefont {Yates}, \citenamefont {Souza},\ and\ \citenamefont
  {Vanderbilt}}]{Wang2006PRB}%
  \BibitemOpen
  \bibfield  {author} {\bibinfo {author} {\bibfnamefont {X.}~\bibnamefont
  {Wang}}, \bibinfo {author} {\bibfnamefont {J.~R.}\ \bibnamefont {Yates}},
  \bibinfo {author} {\bibfnamefont {I.}~\bibnamefont {Souza}}, \ and\ \bibinfo
  {author} {\bibfnamefont {D.}~\bibnamefont {Vanderbilt}},\ }\bibfield  {title}
  {\enquote {\bibinfo {title} {{Ab initio calculation of the anomalous Hall
  conductivity by Wannier interpolation}},}\ }\href {\doibase
  10.1103/PhysRevB.74.195118} {\bibfield  {journal} {\bibinfo  {journal} {Phys.
  Rev. B}\ }\textbf {\bibinfo {volume} {74}},\ \bibinfo {pages} {195118}
  (\bibinfo {year} {2006})}\BibitemShut {NoStop}%
\bibitem [{\citenamefont {Yates}\ \emph {et~al.}(2007)\citenamefont {Yates},
  \citenamefont {Wang}, \citenamefont {Vanderbilt},\ and\ \citenamefont
  {Souza}}]{Yates2007PRB}%
  \BibitemOpen
  \bibfield  {author} {\bibinfo {author} {\bibfnamefont {J.~R.}\ \bibnamefont
  {Yates}}, \bibinfo {author} {\bibfnamefont {X.}~\bibnamefont {Wang}},
  \bibinfo {author} {\bibfnamefont {D.}~\bibnamefont {Vanderbilt}}, \ and\
  \bibinfo {author} {\bibfnamefont {I.}~\bibnamefont {Souza}},\ }\bibfield
  {title} {\enquote {\bibinfo {title} {{Spectral and Fermi surface properties
  from Wannier interpolation}},}\ }\href {\doibase 10.1103/PhysRevB.75.195121}
  {\bibfield  {journal} {\bibinfo  {journal} {Phys. Rev. B}\ }\textbf {\bibinfo
  {volume} {75}},\ \bibinfo {pages} {195121} (\bibinfo {year}
  {2007})}\BibitemShut {NoStop}%
\bibitem [{\citenamefont {Feng}\ \emph {et~al.}(2012)\citenamefont {Feng},
  \citenamefont {Yao}, \citenamefont {Zhu}, \citenamefont {Zhou}, \citenamefont
  {Yao},\ and\ \citenamefont {Xiao}}]{Feng2012PRB}%
  \BibitemOpen
  \bibfield  {author} {\bibinfo {author} {\bibfnamefont {W.}~\bibnamefont
  {Feng}}, \bibinfo {author} {\bibfnamefont {Y.}~\bibnamefont {Yao}}, \bibinfo
  {author} {\bibfnamefont {W.}~\bibnamefont {Zhu}}, \bibinfo {author}
  {\bibfnamefont {J.}~\bibnamefont {Zhou}}, \bibinfo {author} {\bibfnamefont
  {W.}~\bibnamefont {Yao}}, \ and\ \bibinfo {author} {\bibfnamefont
  {D.}~\bibnamefont {Xiao}},\ }\bibfield  {title} {\enquote {\bibinfo {title}
  {{Intrinsic spin Hall effect in monolayers of group-VI dichalcogenides: A
  first-principles study}},}\ }\href {\doibase 10.1103/PhysRevB.86.165108}
  {\bibfield  {journal} {\bibinfo  {journal} {Phys. Rev. B}\ }\textbf {\bibinfo
  {volume} {86}},\ \bibinfo {pages} {165108} (\bibinfo {year}
  {2012})}\BibitemShut {NoStop}%
\bibitem [{\citenamefont {Sun}\ \emph {et~al.}(2016)\citenamefont {Sun},
  \citenamefont {Zhang}, \citenamefont {Felser},\ and\ \citenamefont
  {Yan}}]{Sun2016PRL}%
  \BibitemOpen
  \bibfield  {author} {\bibinfo {author} {\bibfnamefont {Y.}~\bibnamefont
  {Sun}}, \bibinfo {author} {\bibfnamefont {Y.}~\bibnamefont {Zhang}}, \bibinfo
  {author} {\bibfnamefont {C.}~\bibnamefont {Felser}}, \ and\ \bibinfo {author}
  {\bibfnamefont {B.}~\bibnamefont {Yan}},\ }\bibfield  {title} {\enquote
  {\bibinfo {title} {{Strong Intrinsic Spin Hall Effect in the TaAs Family of
  Weyl Semimetals}},}\ }\href {\doibase 10.1103/PhysRevLett.117.146403}
  {\bibfield  {journal} {\bibinfo  {journal} {Phys. Rev. Lett.}\ }\textbf
  {\bibinfo {volume} {117}},\ \bibinfo {pages} {146403} (\bibinfo {year}
  {2016})}\BibitemShut {NoStop}%
\bibitem [{\citenamefont {Zhang}\ \emph {et~al.}(2017)\citenamefont {Zhang},
  \citenamefont {Sun}, \citenamefont {Yang}, \citenamefont
  {\ifmmode~\check{Z}\else \v{Z}\fi{}elezn\'y}, \citenamefont {Parkin},
  \citenamefont {Felser},\ and\ \citenamefont {Yan}}]{Zhang2017PRB}%
  \BibitemOpen
  \bibfield  {author} {\bibinfo {author} {\bibfnamefont {Y.}~\bibnamefont
  {Zhang}}, \bibinfo {author} {\bibfnamefont {Y.}~\bibnamefont {Sun}}, \bibinfo
  {author} {\bibfnamefont {H.}~\bibnamefont {Yang}}, \bibinfo {author}
  {\bibfnamefont {J.}~\bibnamefont {\ifmmode~\check{Z}\else
  \v{Z}\fi{}elezn\'y}}, \bibinfo {author} {\bibfnamefont {S.~P.~P.}\
  \bibnamefont {Parkin}}, \bibinfo {author} {\bibfnamefont {C.}~\bibnamefont
  {Felser}}, \ and\ \bibinfo {author} {\bibfnamefont {B.}~\bibnamefont {Yan}},\
  }\bibfield  {title} {\enquote {\bibinfo {title} {{Strong anisotropic
  anomalous Hall effect and spin {Hall} effect in the chiral antiferromagnetic
  compounds Mn$_{3}$X (X=Ge, Sn, Ga, Ir, Rh, and Pt)}},}\ }\href {\doibase
  10.1103/PhysRevB.95.075128} {\bibfield  {journal} {\bibinfo  {journal} {Phys.
  Rev. B}\ }\textbf {\bibinfo {volume} {95}},\ \bibinfo {pages} {075128}
  (\bibinfo {year} {2017})}\BibitemShut {NoStop}%
\bibitem [{\citenamefont {\ifmmode~\check{Z}\else \v{Z}\fi{}elezn\'y}\ \emph
  {et~al.}(2017)\citenamefont {\ifmmode~\check{Z}\else \v{Z}\fi{}elezn\'y},
  \citenamefont {Zhang}, \citenamefont {Felser},\ and\ \citenamefont
  {Yan}}]{Zelezny2017PRL}%
  \BibitemOpen
  \bibfield  {author} {\bibinfo {author} {\bibfnamefont {J.}~\bibnamefont
  {\ifmmode~\check{Z}\else \v{Z}\fi{}elezn\'y}}, \bibinfo {author}
  {\bibfnamefont {Y.}~\bibnamefont {Zhang}}, \bibinfo {author} {\bibfnamefont
  {C.}~\bibnamefont {Felser}}, \ and\ \bibinfo {author} {\bibfnamefont
  {B.}~\bibnamefont {Yan}},\ }\bibfield  {title} {\enquote {\bibinfo {title}
  {{Spin-Polarized Current in Noncollinear Antiferromagnets}},}\ }\href
  {\doibase 10.1103/PhysRevLett.119.187204} {\bibfield  {journal} {\bibinfo
  {journal} {Phys. Rev. Lett.}\ }\textbf {\bibinfo {volume} {119}},\ \bibinfo
  {pages} {187204} (\bibinfo {year} {2017})}\BibitemShut {NoStop}%
\bibitem [{\citenamefont {Derunova}\ \emph {et~al.}(2019)\citenamefont
  {Derunova}, \citenamefont {Sun}, \citenamefont {Felser}, \citenamefont
  {Parkin}, \citenamefont {Yan},\ and\ \citenamefont
  {Ali}}]{derunova2019sciadv}%
  \BibitemOpen
  \bibfield  {author} {\bibinfo {author} {\bibfnamefont {E.}~\bibnamefont
  {Derunova}}, \bibinfo {author} {\bibfnamefont {Y.}~\bibnamefont {Sun}},
  \bibinfo {author} {\bibfnamefont {C.}~\bibnamefont {Felser}}, \bibinfo
  {author} {\bibfnamefont {S.~S.~P.}\ \bibnamefont {Parkin}}, \bibinfo {author}
  {\bibfnamefont {B.}~\bibnamefont {Yan}}, \ and\ \bibinfo {author}
  {\bibfnamefont {M.~N.}\ \bibnamefont {Ali}},\ }\bibfield  {title} {\enquote
  {\bibinfo {title} {{Giant intrinsic spin Hall effect in W$_3$Ta and other A15
  superconductors}},}\ }\href {\doibase 10.1126/sciadv.aav8575} {\bibfield
  {journal} {\bibinfo  {journal} {Sci. Adv.}\ }\textbf {\bibinfo {volume}
  {5}},\ \bibinfo {pages} {eaav8575} (\bibinfo {year} {2019})}\BibitemShut
  {NoStop}%
\bibitem [{\citenamefont {Zhang}\ \emph {et~al.}(2018)\citenamefont {Zhang},
  \citenamefont {{\v{Z}}elezn{\`y}}, \citenamefont {Sun}, \citenamefont
  {van~den Brink},\ and\ \citenamefont {Yan}}]{Zhang2018NJP}%
  \BibitemOpen
  \bibfield  {author} {\bibinfo {author} {\bibfnamefont {Y.}~\bibnamefont
  {Zhang}}, \bibinfo {author} {\bibfnamefont {J.}~\bibnamefont
  {{\v{Z}}elezn{\`y}}}, \bibinfo {author} {\bibfnamefont {Y.}~\bibnamefont
  {Sun}}, \bibinfo {author} {\bibfnamefont {J.}~\bibnamefont {van~den Brink}},
  \ and\ \bibinfo {author} {\bibfnamefont {B.}~\bibnamefont {Yan}},\ }\bibfield
   {title} {\enquote {\bibinfo {title} {{Spin Hall effect emerging from a
  noncollinear magnetic lattice without spin--orbit coupling}},}\ }\href
  {\doibase 10.1088/1367-2630/aad1eb} {\bibfield  {journal} {\bibinfo
  {journal} {New J. Phys.}\ }\textbf {\bibinfo {volume} {20}},\ \bibinfo
  {pages} {073028} (\bibinfo {year} {2018})}\BibitemShut {NoStop}%
\bibitem [{\citenamefont {Mostofi}\ \emph {et~al.}(2014)\citenamefont
  {Mostofi}, \citenamefont {Yates}, \citenamefont {Pizzi}, \citenamefont {Lee},
  \citenamefont {Souza}, \citenamefont {Vanderbilt},\ and\ \citenamefont
  {Marzari}}]{mostofi2014updated}%
  \BibitemOpen
  \bibfield  {author} {\bibinfo {author} {\bibfnamefont {A.~A.}\ \bibnamefont
  {Mostofi}}, \bibinfo {author} {\bibfnamefont {J.~R.}\ \bibnamefont {Yates}},
  \bibinfo {author} {\bibfnamefont {G.}~\bibnamefont {Pizzi}}, \bibinfo
  {author} {\bibfnamefont {Y.-S.}\ \bibnamefont {Lee}}, \bibinfo {author}
  {\bibfnamefont {I.}~\bibnamefont {Souza}}, \bibinfo {author} {\bibfnamefont
  {D.}~\bibnamefont {Vanderbilt}}, \ and\ \bibinfo {author} {\bibfnamefont
  {N.}~\bibnamefont {Marzari}},\ }\bibfield  {title} {\enquote {\bibinfo
  {title} {{An updated version of wannier90: A tool for obtaining
  maximally-localised Wannier functions}},}\ }\href {\doibase
  10.1016/j.cpc.2014.05.003} {\bibfield  {journal} {\bibinfo  {journal}
  {Comput. Phys. Commun.}\ }\textbf {\bibinfo {volume} {185}},\ \bibinfo
  {pages} {2309} (\bibinfo {year} {2014})}\BibitemShut {NoStop}%
\bibitem [{\citenamefont {Giannozzi}\ \emph {et~al.}(2009)\citenamefont
  {Giannozzi}, \citenamefont {Baroni}, \citenamefont {Bonini}, \citenamefont
  {Calandra}, \citenamefont {Car}, \citenamefont {Cavazzoni}, \citenamefont
  {Ceresoli}, \citenamefont {Chiarotti}, \citenamefont {Cococcioni},
  \citenamefont {Dabo}, \citenamefont {Dal~Corso}, \citenamefont
  {de~Gironcoli}, \citenamefont {Fabris}, \citenamefont {Fratesi},
  \citenamefont {Gebauer}, \citenamefont {Gerstmann}, \citenamefont
  {Gougoussis}, \citenamefont {Kokalj}, \citenamefont {Lazzeri}, \citenamefont
  {Martin-Samos}, \citenamefont {Marzari}, \citenamefont {Mauri}, \citenamefont
  {Mazzarello}, \citenamefont {Paolini}, \citenamefont {Pasquarello},
  \citenamefont {Paulatto}, \citenamefont {Sbraccia}, \citenamefont {Scandolo},
  \citenamefont {Sclauzero}, \citenamefont {Seitsonen}, \citenamefont
  {Smogunov}, \citenamefont {Umari},\ and\ \citenamefont
  {Wentzcovitch}}]{giannozzi2009quantum}%
  \BibitemOpen
  \bibfield  {author} {\bibinfo {author} {\bibfnamefont {P.}~\bibnamefont
  {Giannozzi}}, \bibinfo {author} {\bibfnamefont {S.}~\bibnamefont {Baroni}},
  \bibinfo {author} {\bibfnamefont {N.}~\bibnamefont {Bonini}}, \bibinfo
  {author} {\bibfnamefont {M.}~\bibnamefont {Calandra}}, \bibinfo {author}
  {\bibfnamefont {R.}~\bibnamefont {Car}}, \bibinfo {author} {\bibfnamefont
  {C.}~\bibnamefont {Cavazzoni}}, \bibinfo {author} {\bibfnamefont
  {D.}~\bibnamefont {Ceresoli}}, \bibinfo {author} {\bibfnamefont {G.~L.}\
  \bibnamefont {Chiarotti}}, \bibinfo {author} {\bibfnamefont {M.}~\bibnamefont
  {Cococcioni}}, \bibinfo {author} {\bibfnamefont {I.}~\bibnamefont {Dabo}},
  \bibinfo {author} {\bibfnamefont {A.}~\bibnamefont {Dal~Corso}}, \bibinfo
  {author} {\bibfnamefont {S.}~\bibnamefont {de~Gironcoli}}, \bibinfo {author}
  {\bibfnamefont {S.}~\bibnamefont {Fabris}}, \bibinfo {author} {\bibfnamefont
  {G.}~\bibnamefont {Fratesi}}, \bibinfo {author} {\bibfnamefont
  {R.}~\bibnamefont {Gebauer}}, \bibinfo {author} {\bibfnamefont
  {U.}~\bibnamefont {Gerstmann}}, \bibinfo {author} {\bibfnamefont
  {C.}~\bibnamefont {Gougoussis}}, \bibinfo {author} {\bibfnamefont
  {A.}~\bibnamefont {Kokalj}}, \bibinfo {author} {\bibfnamefont
  {M.}~\bibnamefont {Lazzeri}}, \bibinfo {author} {\bibfnamefont
  {L.}~\bibnamefont {Martin-Samos}}, \bibinfo {author} {\bibfnamefont
  {N.}~\bibnamefont {Marzari}}, \bibinfo {author} {\bibfnamefont
  {F.}~\bibnamefont {Mauri}}, \bibinfo {author} {\bibfnamefont
  {R.}~\bibnamefont {Mazzarello}}, \bibinfo {author} {\bibfnamefont
  {S.}~\bibnamefont {Paolini}}, \bibinfo {author} {\bibfnamefont
  {A.}~\bibnamefont {Pasquarello}}, \bibinfo {author} {\bibfnamefont
  {L.}~\bibnamefont {Paulatto}}, \bibinfo {author} {\bibfnamefont
  {C.}~\bibnamefont {Sbraccia}}, \bibinfo {author} {\bibfnamefont
  {S.}~\bibnamefont {Scandolo}}, \bibinfo {author} {\bibfnamefont
  {G.}~\bibnamefont {Sclauzero}}, \bibinfo {author} {\bibfnamefont {A.~P.}\
  \bibnamefont {Seitsonen}}, \bibinfo {author} {\bibfnamefont {A.}~\bibnamefont
  {Smogunov}}, \bibinfo {author} {\bibfnamefont {P.}~\bibnamefont {Umari}}, \
  and\ \bibinfo {author} {\bibfnamefont {R.M.}\ \bibnamefont {Wentzcovitch}},\
  }\bibfield  {title} {\enquote {\bibinfo {title} {{QUANTUM ESPRESSO: a modular
  and open-source software project for quantum simulations of materials}},}\
  }\href {http://stacks.iop.org/0953-8984/21/i=39/a=395502} {\bibfield
  {journal} {\bibinfo  {journal} {J. Phys.: Condens. Matter}\ }\textbf
  {\bibinfo {volume} {21}},\ \bibinfo {pages} {395502} (\bibinfo {year}
  {2009})}\BibitemShut {NoStop}%
\bibitem [{\citenamefont {Perdew}\ \emph {et~al.}(2008)\citenamefont {Perdew},
  \citenamefont {Ruzsinszky}, \citenamefont {Csonka}, \citenamefont {Vydrov},
  \citenamefont {Scuseria}, \citenamefont {Constantin}, \citenamefont {Zhou},\
  and\ \citenamefont {Burke}}]{PerdewPRL2008}%
  \BibitemOpen
  \bibfield  {author} {\bibinfo {author} {\bibfnamefont {J.~P.}\ \bibnamefont
  {Perdew}}, \bibinfo {author} {\bibfnamefont {A.}~\bibnamefont {Ruzsinszky}},
  \bibinfo {author} {\bibfnamefont {G.~I.}\ \bibnamefont {Csonka}}, \bibinfo
  {author} {\bibfnamefont {O.~A.}\ \bibnamefont {Vydrov}}, \bibinfo {author}
  {\bibfnamefont {G.~E.}\ \bibnamefont {Scuseria}}, \bibinfo {author}
  {\bibfnamefont {L.~A.}\ \bibnamefont {Constantin}}, \bibinfo {author}
  {\bibfnamefont {X.}~\bibnamefont {Zhou}}, \ and\ \bibinfo {author}
  {\bibfnamefont {K.}~\bibnamefont {Burke}},\ }\bibfield  {title} {\enquote
  {\bibinfo {title} {{Restoring the Density-Gradient Expansion for Exchange in
  Solids and Surfaces}},}\ }\href {\doibase 10.1103/PhysRevLett.100.136406}
  {\bibfield  {journal} {\bibinfo  {journal} {Phys. Rev. Lett.}\ }\textbf
  {\bibinfo {volume} {100}},\ \bibinfo {pages} {136406} (\bibinfo {year}
  {2008})}\BibitemShut {NoStop}%
\bibitem [{\citenamefont {Qiao}\ \emph {et~al.}(2018)\citenamefont {Qiao},
  \citenamefont {Zhou}, \citenamefont {Yuan},\ and\ \citenamefont
  {Zhao}}]{qiao2018PRB}%
  \BibitemOpen
  \bibfield  {author} {\bibinfo {author} {\bibfnamefont {J.}~\bibnamefont
  {Qiao}}, \bibinfo {author} {\bibfnamefont {J.}~\bibnamefont {Zhou}}, \bibinfo
  {author} {\bibfnamefont {Z.}~\bibnamefont {Yuan}}, \ and\ \bibinfo {author}
  {\bibfnamefont {W.}~\bibnamefont {Zhao}},\ }\bibfield  {title} {\enquote
  {\bibinfo {title} {{Calculation of intrinsic spin Hall conductivity by
  Wannier interpolation}},}\ }\href {\doibase 10.1103/PhysRevB.98.214402}
  {\bibfield  {journal} {\bibinfo  {journal} {Phys. Rev. B}\ }\textbf {\bibinfo
  {volume} {98}},\ \bibinfo {pages} {214402} (\bibinfo {year}
  {2018})}\BibitemShut {NoStop}%
\end{thebibliography}%

\end{document}